\documentclass[final,5p,times,twocolumn]{elsarticle}

\usepackage{amssymb}
\usepackage{lipsum}

\usepackage{hyperref}       
\usepackage{url}            
\usepackage[printonlyused,nohyperlinks]{acronym}
\usepackage{siunitx}
\usepackage{booktabs}       
\usepackage{amsfonts}       
\usepackage{amssymb}
\usepackage{pifont}
\usepackage[export]{adjustbox}
\usepackage{orcidlink}
\usepackage[switch]{lineno}
\usepackage{setspace}

\newcommand{\cmark}{\ding{51}}%
\newcommand{\xmark}{\ding{55}}%

\newcommand{\meanSD}[2]{\mbox{$(M\!=\!#1,~SD\!=\!#2)$}}
\newcommand{\pValue}[1]{\mbox{$p\!=\!#1$}}
\newcommand{\pValuePL}{\mbox{$p\!<\!.001$}}

\newcommand{\ChiSquarePL}[3]{\mbox{$(\chi^2(#1)\!=\!#2,~p\!<\!.001)$}}

\newcommand{\studentT}[3]{\mbox{$t(#1)\!=\!#2,~p\!=\!#3$}}
\newcommand{\yuenT}[3]{\mbox{$\mathrm{Yuen's-}t(#1)\!=\!#2,~p\!=\!#3$}}
\newcommand{\yuenTPL}[3]{\mbox{$\mathrm{Yuen's-}t(#1)\!=\!#2,~p\!<\!.001$}}

\newcommand{\mannWhitney}[2]{\mbox{$U\!=\!#1,~p\!=\!#2$}}

\newcommand{\wilcoxon}[2]{\mbox{$(W\!=\!#1,~p\!=\!#2)$}}

\newcommand{\trafficPreview}{\rho_T}

\newcommand{\trafficGradient}{\nabla_T}
\newcommand{\trafficLateralDistance}{d_{T,0}}

\newcommand{\pLevelOne}{$^*$}
\newcommand{\pLevelTwo}{$^{**}$}
\newcommand{\pLevelThree}{$^{***}$}

\newcommand{\clineSpacing}{\noalign{\vspace{0.05cm}}}

\newcommand{\tableVSpacing}{0.1cm}

\journal{~}

\begin{document}

\begin{frontmatter}



\title{Exploring the Influence of Driving Context on Lateral Driving Style Preferences: A Simulator-Based Study}


\author[tu,ifm,c]{Johann Haselberger\,\orcidlink{0000-0002-2458-3461}}
\author[tu,ifm]{Maximilian Böhle\,\orcidlink{0000-0002-2919-000X}}
\author[ifm]{Bernhard Schick\,\orcidlink{0000-0001-5567-3913}}
\author[tu]{Steffen Müller}
\affiliation[tu]{
    organization={Department of Automotive Engineering, Technical University Berlin},
    addressline={Straße des 17. Juni 135}, 
    city={Berlin},
    postcode={10623}, 
    country={Germany}
}
\affiliation[ifm]{
    organization={Institute for Driver Assistance and Connected Mobility, University of Applied Science Kempten},
    addressline={Bahnhofstraße 61}, 
    city={Kempten},
    postcode={87435}, 
    country={Germany}
}
\affiliation[c]{Corresponding author; johann.haselberger@hs-kempten.de}

\begin{abstract}
Technological advancements focus on developing comfortable and acceptable driving characteristics in autonomous vehicles.
Present driving functions predominantly possess predefined parameters, and there is no universally accepted driving style for autonomous vehicles.
While driving may be technically safe and the likelihood of road accidents is reduced, passengers may still feel insecure due to a mismatch in driving styles between the human and the autonomous system.
Incorporating driving style preferences into automated vehicles enhances acceptance, reduces uncertainty, and poses the opportunity to expedite their adoption.
Despite the increased research focus on driving styles, there remains a need for comprehensive studies investigating how variations in the driving context impact the assessment of automated driving functions.
Therefore, this work evaluates lateral driving style preferences for autonomous vehicles on rural roads, considering different weather and traffic situations.
A controlled study was conducted with a variety of German participants utilizing a high-fidelity driving simulator.
The subjects experienced four different driving styles, including mimicking of their own driving behavior under two weather conditions.
A notable preference for a more passive driving style became evident based on statistical analyses of participants' responses during and after the drives.
This study could not confirm the hypothesis that subjects prefer to be driven by mimicking their own driving behavior.
Furthermore, the study illustrated that weather conditions and oncoming traffic substantially influence the perceived comfort during autonomous rides. 
The gathered dataset is openly accessible at \href{https://www.kaggle.com/datasets/jhaselberger/idcld-subject-study-on-driving-style-preferences}{\textbf{https://www.kaggle.com/datasets/jhaselberger/idcld-subject-study-on-driving-style-preferences}}.
\end{abstract}




\end{frontmatter}


\section{Introduction}

The shift to fully autonomous driving is an ongoing, lengthy process \cite{roos2020technologie}.
Despite long-standing research on \ac{avs} \cite{Cox1991}, active development continues, driven by rapid advancements in hardware and software technology \cite{eltawab2020, Yan2020}.
The widespread use of \ac{avs} is expected to reduce accidents, traffic, and carbon emissions \cite{ertrac2019connected}.
As technology evolves, the focus moves from mere feasibility to realizing comfortable and acceptable driving characteristics \cite{bellem2018comfort}.
Systems should meet not only their stated promises but also the expectations of users \cite{drewitz2020towards}.
Drivers anticipate that automated driving systems will function consistently with their own actions \cite{basu2017you,peng2022drivers,hartwich2018driving}.
A comfortable, safe ride experience is vital for the adoption of \ac{avs} \cite{bellem2018comfort,lee2021assessing,voss2018investigation,xiao2010comprehensive,bengtsson2001adaptive}.
Unfortunately, drivers hesitate to use autonomous driving functions due to their limited trust and acceptance of the driving styles exhibited by AVs \cite{ma2021drivers}.
While driving may be technically safe, the passenger may still experience insecurity due to differences in driving styles between the human and the autonomous system \cite{bolduc2019multimodel}.
Hence, driving styles play a crucial role in fostering trust and acceptance of \ac{avs} \cite{ekman2019exploring,strauch2019real,carsten2019can,ramm2014first}.
The driving style of \ac{avs} should be reliable and familiar and avoid sudden surprise behaviors to enhance acceptance, satisfaction, and perceived safety \cite{carsten2019can,ramm2014first}.
However, there is a lack of understanding regarding peoples' preferences for being driven in a highly automated vehicle \cite{radlmayr2015literaturanalyse,gasser2013herausforderung,siebert2013discomfort,rossner2019diskomfort, rossner2022also}.
Different user groups prefer distinct autonomous driving styles \cite{peng2022drivers}.
Up to now, there is no comprehensive and standardized definition of the term "driving style" \cite{itkonen2020characterisation,chu2020self,chen2021driving}.
However, definitions share the common idea that driving style comprises a set of driving habits that a driver forms and refines over time with increasing driving experience \cite{elander1993behavioral, lajunen2011self,sagberg2015review,kleisen2011relationship,tement2022assessment}.
Driving style is considered a relatively stable driver trait \cite{saad2004behavioural,sagberg2015review,he2022individual}, characterized by resistance to significant changes over a short time period, as it reflects deeply ingrained driving habits \cite{martinez2017driving,tement2022assessment}.
Driving style is distinct from driving skill, the information-processing and psychomotor abilities of a driver, as the underlying habits do not necessarily evolve with increasing experience \cite{tement2022assessment}.

The inflexible driving style of \ac{avs} that disregards user preferences not only creates discomfort but also impacts system acceptance and consistent usage by drivers and other road users \cite{price2016psychophysics,siebert2013discomfort,delmas2022effects,dillen2020keep}.
It is assumed that factors influencing perceived safety in manual driving also apply to perceived safety during automated driving \cite{rossner2020care}.
The dominant presumption, albeit not explicitly stated, is that drivers prefer a driving style that mirrors their own \cite{hasenjager2019survey,festner2016einfluss,griesche2016should,bolduc2019multimodel,sun2020exploring,hartwich2015drive,rossner2022also,dettmann2021comfort}.
This assumption suggests that trajectories derived from a human driving style have significant potential to improve perceived safety \cite{bellem2017can, lex2017objektive, rossner2018drive, rossner2019you} as humans are skeptical of autonomous vehicles taking over control, often placing more trust in their own skills than in the vehicle itself \cite{schoettle2014survey}.
Users' trust was found to be higher for \ac{avs} that closely match their preferences \cite{natarajan2022toward}.
In addition, trust is enhanced when there is alignment between an agent's capability and the given situation \cite{robert2009individual,petersen2019situational}.
Situational awareness involves perceiving the vehicle's state and the traffic environment, comprehending the current situation, and predicting future developments \cite{stanton2001situational}.
When users perceive a mismatch between their assessment of a situation and the behavior of the autonomous machine, they tend to override it \cite{riley1989general}.
According to \cite{chen2021semi}, the driving context can be categorized into two types: dynamic conditions (e.g., traffic light states or changing weather) and static conditions (e.g., road geometry or lane distribution).

The driving context has a significant influence on driving behavior \cite{dong2016characterizing, shouno2018deep, han2019statistical,ghasemzadeh2018utilizing,constantinescu2010driving,chen2019graphical,chen2021driving,hamdar2016weather}.
The way individuals respond to diverse driving contexts constitutes a significant aspect of the driving style \cite{chen2019driving} that, however, has often been neglected in previous works \cite{bejani2018context}.
While there is strong evidence that weather influences driving behavior \cite{ahmed2018impacts,kilpelainen2007effects,rahman2012analysis,faria2020assessing}, the degree of variation in driving behavior due to changing weather conditions varies among individual drivers \cite{hamada2016modeling}.
Poor visibility, such as fog, has been observed to impact driving behavior, with drivers increasing their following distances, as reported in studies like \cite{hamdar2016weather} and \cite{van1998strategic}. Conversely, in \cite{evans2004traffic} instances were reported where drivers chose to maintain a shorter distance to the leading vehicle due to concerns about losing a reference point.
Besides the effects introduced by weather conditions, traffic also plays a crucial role in shaping the perceived safety during automated driving.
Particularly when faced with oncoming traffic, drivers tend to deviate from the lane center \cite{rossner2020care,bellem2017can,lex2017objektive,rossner2022also,schlag2015auswirkungen,rosey2009impact,triggs1997effect}.
Especially oncoming trucks cause stronger responses \cite{dijksterhuis2012adaptive,mecheri2017effects,schlag2015auswirkungen, rossner2022also, spacek2005track,rosey2009impact,rasanen2005effects} as they notably reduce the sense of safety.
Since drivers are unable to react to oncoming traffic, the \ac{avs} must adjust their trajectories to enhance both perceived safety and comfort \cite{rossner2020care}.
This more responsive approach to negotiating curves differs from the majority of existing autonomous driving policies that typically adhere closely to the lane center.
In practical, real-world driving situations, strict lane-center tracking is not commonly observed \cite{gordon2014modeling,rossner2020care,haselberger2023self}.
Based on convenience, drivers tend to choose a shorter trajectory when navigating curves \cite{ding2014driver}.
The fundamental human lane-keeping process lacks explicit trajectory planning, with drivers instead estimating an expected lateral driving zone, which can be treated as an acceptable safety area \cite{ding2014driver}.

Given the limited number of studies exploring driving style preferences in automated driving and their diverse focus on various driving situations and maneuvers, a comprehensive understanding of users' desired experience in AVs remains elusive \cite{vasile2023influences}.
Therefore, this study evaluates lateral driving style preferences for \ac{avs} on rural roads, considering different weather and traffic situations.
In particular, there is limited work in understanding how adverse weather conditions impact the assessment of automated driving functions.
While earlier studies predominantly emphasize the longitudinal aspects of driving style, it is essential to note that the lateral component, particularly on rural roads, plays a crucial role in ensuring road safety and thus poses a crucial factor in subjects' comfort and trust ratings.
Our contributions can be summarized to:
\begin{enumerate}
    \item Conduction of a controlled driving study utilizing a high-fidelity driving simulator.
    \item Introducing a novel reactive driving behavior model that can emulate human-like curve negotiation while responding to oncoming traffic.
    \item Statistical evaluation of subjects' \acl{tia}, \acl{arca} and Relaxation Level responses combined with the \acl{mdsi} self-assessments.
    \item Evaluation of the agreement of subjective driving style self-assessments with the collected subjective driving style ratings.
    \item Publicly accessible provision of the dataset including the anonymized socio-demographics and questionnaire responses.
\end{enumerate}
\section{Related Work}

Regarding the research focus, previous works predominantly concentrated on driving style preferences, primarily considering longitudinal aspects on highways and more artificial test tracks.
An overview of related driving studies and the considered distinctive attributes is given in \autoref{tab:relatedWork}.
There is a shortcoming in the in-depth analysis of differences in driving style preferences regarding lateral driving behavior on rural roads under different driving situations.
Compared to highways, lane-keeping tasks on rural roads are more challenging.
In manual driving, run-off-the-road crashes and near-crash incidents can be ascribed to drivers' inadequate lane-keeping performance \cite{ghasemzadeh2018utilizing, ghasemzadeh2017drivers}.
This raises the question of whether participants in this setting are inclined to accept a sportier driving style from the autonomous vehicle or lean more towards a passive one.
While numerous references acknowledge that weather conditions influence driving style \cite{ahmed2018impacts,kilpelainen2007effects,rahman2012analysis,faria2020assessing,evans2004traffic,van1998strategic,hamdar2016weather}, limited research specifically investigates the impact on perceived comfort and passenger preferences when driven by automated driving functions.
In conjunction with various traffic scenarios, the driving context presents possibilities or imposes limitations on action selection \cite{sagberg2015review}.
While manually driving, external factors hinder drivers from attaining theoretically ideal driving behaviors \cite{magana2018method}.
Therefore, it is crucial to consider these external influences when assessing automated driving functions.


\begingroup

\setlength{\tabcolsep}{2pt} 

\begin{table*}[t]
    \caption{Related subject studies in chronological order. N represents the number of subjects, the country codes follow the ISO 3166-1 encoding. \cmark, (\cmark), and \xmark~indicate whether a requirement is met, partially met or not fulfilled. (+/++/+++) are denoting, as objectively as possible, the fidelity of the used simulator: + for more simple static simulators, ++ for fixed-based driving simulators with a fully equipped interior, and +++ for incorporating a motion system and realistic road models.}
    \centering
    \scriptsize

    \begin{tabular}{lccclccclcccclcclccclcccccccccc}
        \toprule
        \textbf{}                 & \textbf{}     & \textbf{}  & \textbf{}        & \textbf{} & \multicolumn{3}{c}{\textbf{Domain}}                               & \textbf{} & \multicolumn{4}{c}{\textbf{Context}}                                    & \textbf{} & \multicolumn{2}{c}{\textbf{Focus}}       & \textbf{} & \multicolumn{3}{c}{\textbf{Variations}}                               & \textbf{} & \multicolumn{7}{c}{\textbf{Driving Style   Representation}}                                                                                                                                                                      & \textbf{}                                 & \textbf{}                                                                                                                                                                                      & \textbf{}                  \\ \cline{6-8} \cline{10-13} \cline{15-16} \cline{18-20} \cline{22-28}
        \textbf{Ref.}             & \textbf{Year} & \textbf{N} & \rotatebox{90}{\textbf{Country}} & \textbf{} & \rotatebox{90}{\textbf{Real}} & \rotatebox{90}{\textbf{Simulation}} & \rotatebox{90}{\textbf{Simulator Fidelity}} & \textbf{} & \rotatebox{90}{\textbf{City}} & \rotatebox{90}{\textbf{Rural}} & \rotatebox{90}{\textbf{Highway}} & \rotatebox{90}{\textbf{Test Track}} & \textbf{} & \rotatebox{90}{\textbf{Lateral}} & \rotatebox{90}{\textbf{Longitudinal}} & \textbf{} & \rotatebox{90}{\textbf{Static Traffic}} & \rotatebox{90}{\textbf{Dynamic Traffic}} & \rotatebox{90}{\textbf{Weather}} & \textbf{} & \rotatebox{90}{\textbf{Own Replay}} & \rotatebox{90}{\textbf{Replay of Others}} & \rotatebox{90}{\textbf{Driving Style Indicators}} & \rotatebox{90}{\textbf{\iffalse Predefinded \fi Acceleration Profiles}} & \rotatebox{90}{\textbf{\iffalse Predefined \fi Trajectories}} & \rotatebox{90}{\textbf{Machine Learning}} & \rotatebox{90}{\textbf{Tuned by Measurements~}} & \textbf{} & \rotatebox{90}{\textbf{On-Drive \iffalse  Comfort/Trust \fi Assessment}}                                                                                                                                                                          & \rotatebox{90}{\textbf{Data Availability}} \\
        \midrule
        \cite{scherer2016will}           & 2016          & 20         & DE               &           & \cmark             &                     &                             &           &               &                &                  & \cmark                   &           &                  &\cmark                    &           &                         &                          &                  &           &\cmark                  &                           &                                   &\cmark                                         &                                  &                           &                                &           &\cmark                                                                                                                                                                                                                                   & \xmark                     \\
        \cite{basu2017you}               & 2017          & 15         & US               &           &               & \cmark                   & +                         &           & \cmark             &                & \cmark                & \cmark                   &           & (\cmark)              &\cmark                    &           &                         &\cmark                       &                  &           &\cmark                  &                           &\cmark                                &                                            &                                  &                           &                                &           &                                                                                                                                                                                                                                      & \xmark                     \\
        \cite{bellem2018comfort}         & 2018          & 72         & DE               &           &               & \cmark                   & +++                       &           &               &                & \cmark                &                     &           & (\cmark)              &\cmark                    &           &                         & (\cmark)                      &                  &           &                     &                           &\cmark                                &\cmark                                         & (\cmark)                              &                           &                                &           &                                                                                                                                                                                                               & \xmark                     \\
        \cite{hartwich2018driving}       & 2018          & 46         & DE               &           &               & \cmark                   & ++                        &           &               & \cmark              & \cmark                &                     &           &                  &\cmark                    &           &                         & (\cmark)                      &                  &           &\cmark                  &\cmark                        &                                   &                                            &                                  &                           &                                &           &\cmark                                                                                                                                                                                      & \xmark                     \\
        \cite{rossner2019diskomfort}     & 2019          & 46         & DE               &           &               & \cmark                   & ++                        &           &               &                & \cmark                &                     &           &                  &\cmark                    &           &                         & (\cmark)                      &                  &           &                     &                           &                                   &\cmark                                         &\cmark                               &                           &                                &           &\cmark                                                                                                                                                                                                                                  & \xmark                     \\
        \cite{siebert2019speed}          & 2019          & 39         & DE               &           &               & \cmark                   & ++                        &           & \cmark             & \cmark              & \cmark                & \cmark                   &           &                  &\cmark                    &           &                         &\cmark                       &\cmark               &           &                     &                           &\cmark                                &                                            &                                  &                           &                                &           &\cmark                                                                                                                                                                                                                                     & \xmark                     \\
        \cite{ekman2019exploring}        & 2019          & 18         & SE               &           & \cmark             &                     &                             &           & (\cmark)           & (\cmark)            &                  & \cmark                   &           & (\cmark)              &\cmark                    &           &                         &\cmark                       &                  &           &                     & (\cmark)                       &                                   &                                            &                                  &                           &                                &           &\cmark                                                                                                                                                                                                      & \xmark                     \\
        \cite{oliveira2019driving}       & 2019          & 43         & UK               &           & ~\cmark$^*$          &                     &                             &           & (\cmark)          &                &                  & \cmark                  &           &       \cmark           &         \cmark              &           &                         &\cmark                       &                  &           &                     &                           &                                   &                                            &\cmark                               &                           &                                &           &                                                                                                                                                                                                           & \xmark                     \\
        \cite{rossner2020care}           & 2020          & 30         & DE               &           &               & \cmark                   & ++                        &           &               & (\cmark)            &                  & \cmark                   &           &\cmark               &                       &           &                         &\cmark                       &                  &           &                     &                           &\cmark                                &                                            &                                  &                           &                                &           &\cmark                                                                                                                                                   & \xmark                     \\
        \cite{sun2020exploring}          & 2020          & 36         & CN               &           &               & \cmark                   & ++                        &           &               & (\cmark)            &                  & \cmark                   &           &                  &\cmark                    &           &                         &\cmark                       &                  &           &                     &                           &\cmark                                &                                            &                                  &                           &\cmark                             &           &                                                                                                                                                                                                     & \xmark                     \\
        \cite{dillen2020keep}            & 2020          & 20         & CA               &           & \cmark             &                     &                             &           &               &                &                  & \cmark                   &           &\cmark               &\cmark                    &           &\cmark                      & (\cmark)                      &                  &           &                     &                           &\cmark                                &                                            &                                  &                           &\cmark                             &           &\cmark                                                                                                                                                                                                                          & \cmark                     \\
        \cite{ma2021drivers}             & 2021          & 32         & US               &           &               & \cmark                   & +                         &           & \cmark            &                &                  &                     &           & (\cmark)              &\cmark                    &           &                         &\cmark                       &                  &           &                     &                           &\cmark                                &                                            &                                  &                           & (\cmark)                            &           &\cmark                                        & \xmark                     \\
        \cite{hajiseyedjavadi2022effect} & 2022          & 24         & UK               &           &               & \cmark                   & +++                       &           & \cmark             & \cmark              &                  &                     &           &\cmark               &\cmark                    &           &\cmark                      &                          &                  &           &\cmark                  &                           &\cmark                                &                                            &                                  &                           &\cmark                             &           &\cmark                                                                                                                                        & \xmark                     \\
        \cite{peng2022drivers}           & 2022          & 24         & UK               &           &               & \cmark                   & +++                       &           & \cmark             & \cmark              &                  &                     &           &\cmark               &\cmark                    &           &                         &                          &                  &           &                     &\cmark                        &                                   &                                            &                                  &\cmark                        &\cmark                             &           &\cmark                                                                                                                                                                              & \xmark                     \\
        \cite{ossig2022tactical}         & 2022          & 60         & DE               &           & \cmark             &                     &                             &           &               &                & \cmark                &                     &           &\cmark               &                       &           &                         &\cmark                       &                  &           &                     &                           &\cmark                                &                                            &                                  &                           &\cmark                             &           &                                                                                                                                          & \xmark                     \\
        \cite{wang2022classification}    & 2022          & 12         & JP               &           &               & \cmark                   & +++                       &           &               &                & \cmark                &                     &           &\cmark               & (\cmark)                   &           &                         &\cmark                       &                  &           &                     &                           &\cmark                                &                                            &                                  &                           &\cmark                             &           &                                                                                                                                                                                                                                      & \xmark                     \\
        \cite{kamaraj2023comparing}      & 2023          & 24         & US               &           &               & \cmark                   & +                         &           & \cmark             &                &                  &                     &           &                  &\cmark                    &           &                         &\cmark                       &                  &           &                     &                           &\cmark                                &                                            &                                  &                           &\cmark                             &           &                                                                                                                                                                                                                                 & \xmark                     \\
        \cite{vasile2023influences}      & 2023          & 42         & DE               &           & \cmark             &                     &                             &           &               &                & \cmark                &                     &           &                  &\cmark                    &           &                         &\cmark                       &                  &           &                     &                           &\cmark                                &                                            &                                  &                           &\cmark                             &           &\cmark                                                                                                   & \xmark                     \\
        \cite{schrum2023maveric}         & 2023          & 24         & US               &           &              & \cmark                    &     +++                        &           &               &                & \cmark                &                     &           &     (\cmark)             &\cmark                    &           &                         &\cmark                       &                  &           &                     &                           &                                &                                            &                                &         \cmark                    &\cmark                             &           &                                                                                                   & \xmark                     \\
        \cite{delmas2024personalizing}   & 2024          & 52         & FR               &           &               & \cmark                   & ++                        &           &               &                     & \cmark           &                     &           & (\cmark)            & \cmark                     &           & \cmark                  &                             &                     &           &                        &                           & \cmark                               &                                            &                                  &                           & \cmark                            &           &                                                                                                                                                                                                                        & \cmark         \\
        \textbf{ours}                    & 2024          & 42         & DE               &           &               & \cmark                   & +++                       &           &               & \cmark              &                  &                     &           &\cmark               & (\cmark)                   &           &                         &\cmark                       &\cmark               &           &\cmark                  &                           &\cmark                                &                                            &                                  &                           &\cmark                             &           &\cmark                                                                                                                                                                                                                  & \cmark         \\
        \bottomrule
        \multicolumn{31}{l}{\footnotesize{$^*$ utilizes autonomous people movers instead of road vehicles}} \\                
        \end{tabular}

\label{tab:relatedWork}
\end{table*}

\endgroup


Considering the experimental setup conducted in a real traffic environment, real-world studies exhibit strong external validity.
However, they are limited in terms of internal validity due to the varying traffic conditions experienced by each participant.
Only the experiments in \cite{vasile2023influences} and \cite{ossig2022tactical} are conducted on actual highways in real-world conditions without limiting the traffic situation to a dedicated test track.
However, both studies lack the incorporation of weather conditions, which would not have been feasible without significant additional effort, such as multiple measurement drives on different days.
Employing driving simulators is a viable and widely used option to ensure consistent environmental conditions for each study participant.
Moreover, with assured repeatability, there is the flexibility to systematically modify elements like the road, traffic, weather, and other factors in a simple and replicable manner with low inherent risks and reduced costs \cite{hamdar2016weather,qi2019vehicle,mohammadnazar2021classifying}.
There is substantiated evidence that a driving simulator serves as a valid tool for analyzing driving behavior, as there is a good agreement between the behavior in a driving simulator and real-world driving \cite{van2018relation, changbin2015driving, zhao2014partial, meuleners2015validation,schluter2021identifikation}.
The varying fidelity of the used driving simulators, however, affects the perception of the driver and, consequently, the observed effects.
A comparison of the related works indicates that, in certain instances, relatively simple static simulators are employed \cite{basu2017you,ma2021drivers,kamaraj2023comparing}, and frequently, real vehicle cabins with a fully equipped interior are utilized \cite{hartwich2018driving,rossner2019you,siebert2019speed,rossner2020care,sun2020exploring,delmas2024personalizing}, albeit not in conjunction with a motion system.
Without incorporating a motion system, the absence of proprioceptive feedback in a driving simulator diminishes the realism of the experience and increases the level of difficulty, leading to undesired high velocities, accelerations, and turning issues \cite{van2018relation}.
Improving immersion in the simulator additionally involves ensuring that the vehicle looks and behaves like a real car and that the behavior of traffic actors appears realistic \cite{helman2015validation}.
Therefore, when evaluating driving style differences, it is crucial to utilize a motion system in combination with realistic road and traffic models like in \cite{bellem2018comfort,hajiseyedjavadi2022effect,peng2022drivers,wang2022classification,schrum2023maveric}.

The most commonly utilized parameters in related studies to characterize distinct driving styles include longitudinal and lateral accelerations \cite{dillen2020keep, ma2021drivers, wang2022classification, kamaraj2023comparing, bellem2018comfort} as well as jerk values \cite{sun2020exploring, bellem2018comfort}.
In the context of driving behavior analysis, acceleration and jerk values have consistently proven effective for the classification of driving styles and, as a result, are frequently employed in other works \cite{van2018relation, murphey2009driver, rath2019lane, feng2018driving, martinez2017driving, chu2017curve, bellem2016objective, bellem2017can, vilaca2017systematic, kanarachos2018smartphones, freyer2007ein}.
Regarding longitudinal driving behavior, comparative studies vary the driving speed \cite{basu2017you,sun2020exploring,ma2021drivers,hajiseyedjavadi2022effect,vasile2023influences, delmas2024personalizing} in combination with different headway distances and times \cite{basu2017you,dillen2020keep,wang2022classification,siebert2019speed,ma2021drivers,vasile2023influences}.
The lateral driving behavior is characterized by the distance to the centerline \cite{rossner2020care,hajiseyedjavadi2022effect} and the frequency and duration of lane changes \cite{ossig2022tactical,wang2022classification}.
Besides these relatively simple parameters, replays of the driver's own driving behavior, whether exclusively \cite{scherer2016will,basu2017you,hajiseyedjavadi2022effect} or in conjunction with replays from other study participants \cite{hartwich2018driving,ekman2019exploring,peng2022drivers}, are frequently employed.
While specific parameters may only cover a certain part of the driving style spectrum, replays reproduce all aspects of human driving style.
However, the recorded trajectories are not easily modifiable, posing a challenge to isolate and analyze particular effects.
In addition to these two approaches, pre-configured acceleration profiles \cite{scherer2016will,bellem2018comfort,rossner2019diskomfort} and trajectories \cite{bellem2018comfort,rossner2019diskomfort,oliveira2019driving} are occasionally employed.
A completely different approach is taken by data-driven methods that utilize machine learning to create a driving-style model based on existing data.
In \cite{peng2022drivers}, a \ac{rcnn} was employed to emulate human driving behavior, predicting future yaw rate and speed demands.
The study presented in \cite{schrum2023maveric} introduces a framework where a high-level model is trained to adjust low-level controllers, enabling the learning of a personalized driving style embedding and the modulation of aggressiveness while preserving overall driving style characteristics.

Regardless of the representation of driving style, it is evident that in recent studies, the adjustable parameters for forming different driving style configurations are determined not by heuristics but by incorporating actual experiments or dedicated subject studies.
Almost every of the compared related studies assesses driving style preferences using questionnaires after the drive.
In this context, the \ac{mdsi} \cite{taubman2004multidimensional} and \ac{tia} \cite{korber2019theoretical} are the most frequently utilized ones.
Other questionnaires such as the Thrill and Adventure Seeking subscale of the Sensation Seeking Scale V \cite{zuckerman1978sensation}, Locus of Control \cite{rotter1966generalized}, Trust Questionnaire \cite{jian1998towards}, \ac{csai} \cite{martens1990development}, \ac{ads} \cite{krahe2002predicting}, Checklist for Trust between People and Automation \cite{jian2000foundations}, Propensity to Trust Questionnaire \cite{sinha2008role}, System Acceptance Questionnaire \cite{van1997simple}, Traffic Locus of Control (T-LOC) \cite{ozkan2005multidimensional}, Driving Style Questionnaire (DSQ) \cite{elander1993behavioral}, Arnett Inventory of Sensation Seeking \cite{arnett1994sensation}, and the UX questionnaire \cite{bartneck2009measurement} are also occasionally used.
Self-reports have proven to be reliable as well as valid \cite{taubman2016value,lajunen2003can}, and the MDSI was found to be a good indicator of driving behavior \cite{kaye2018comparison,van2018relation,taubman2016value,haselberger2023self}.
Nevertheless, findings derived from self-reports may not always be optimal for objective classifications \cite{kovaceva2020identification}, as they may be subject to a bias toward providing socially desirable responses \cite{crowne1960new, lajunen1997speed, bakhshi2022evaluating, tement2022assessment}.
Specialized hand controllers are typically used to assess comfort during the drive \cite{scherer2016will,rossner2019diskomfort,hartwich2018driving,rossner2020care,siebert2019speed}, allowing participants to express their feelings continuously.
Moreover, there are on-drive evaluations based on verbal feedback or questionnaires, typically initiated through prompts for assessment by a research assistant or an audio signal \cite{ekman2019exploring,peng2022drivers,vasile2023influences}.

Subject studies typically entail a substantial time commitment, even when conducted within a driving simulator.
This underscores the value of the collected data and the insights derived from such studies.
To the best of the authors' knowledge, only datasets from \cite{dillen2020keep} and \cite{delmas2024personalizing} have been publicly released.
To facilitate additional research encompassing lateral and longitudinal driving aspects, we have made our dataset, including anonymized socio-demographics, questionnaire responses, simulator measurements, and associated labels, openly accessible.

\section{Method}

\subsection{Participants}

The experiment involved \num{42} participants (\num{13} female, \num{29} male).
However, ten subjects had to terminate prematurely due to motion sickness.
Therefore, all subsequent statistics and evaluations refer to the remaining \num{32} subjects fully participating in the study.
Participants were required to hold a driver's license for a minimum of two years.
Subjects held their driver's licenses between four and \num{48} years \meanSD{13.8}{10.7}.
The age covered a range from \num{21} to \num{61} years, and the average age was $29.10 \pm 7.61$ years.
To the question of having ever driven in a driving simulator, $50\%$ of the subjects answered never, $20.6\%$ rarely, $17.6\%$ occasionally, and $11.8\%$ regularly.
Participation in the two-fold survey study was purely voluntary, without any compensation or reward.
All personal data were anonymized.
\subsection{Instruments}
At various time points throughout the study, participants were requested to fill out questionnaires: before the driving experiment, during autonomous driving experiences, after each individual drive, and upon completion of the entire study.
The initial inventory was provided online and comprises general inquiries about the participants, encompassing information such as age, gender, annual mileage, frequency of vehicle usage, possession of a driver's license, preferred average highway speed, engine power of their personal vehicle, and a single self-assessment item regarding their driving style on highways and rural roads.
For a more sophisticated estimation of the subjects' driving style, the \ac{mdsi} \cite{taubman2004multidimensional} was utilized.
The \ac{mdsi} incorporates items from various other questionnaires, including the \ac{dbi} \cite{gulian1989dimensions}, the \ac{dbq} \cite{reason1990errors}, and the \ac{dsq} \cite{french1993decision}.
The inventory consists of 44 items and encompasses four driving style aspects: patient and careful, angry and hostile, reckless and careless, and anxious.
In previous studies, self-reports have demonstrated reliability and validity \cite{taubman2016value, lajunen2003can}, and the MDSI has been identified as a robust indicator of driving behavior \cite{kaye2018comparison, van2018relation, taubman2016value}.
Translations and validations of the scale have been conducted for use in diverse cultural contexts \cite{freuli2020cross, poo2013reliability, wang2018effect, long2019reliability, holman2015romanian, van2015measuring, padilla2020adaptation}.
This study uses the German version of the \ac{mdsi} proposed in \cite{haselberger2023self}.
Participants scored their agreement on each MDSI item using a 6-point Likert scale, ranging from 1 (not at all) to 6 (very much).

In order to query the subjects' overall experience, feelings, and trust, both the \ac{arca} \cite{marberger2022questionnaire} and the \ac{tia} \cite{korber2019theoretical} questionnaire are asked after each individual ride.
Moreover, the subjects needed to provide an estimate of the presented driving style, including the possible choices "passive", "like on rails", "similar to mine", and "sportive".
The \ac{arca} pertains to aspects of ride comfort in automated vehicles that are associated with the design of the vehicle's motion.
It purposefully does not cover aspects of ride comfort influenced by factors like vehicle suspension, seat ergonomics, or interior layout.
Out of the original \num{19} items, this study utilizes a reduced selection of five items, which are more in line with the overall research aim and driving experiment.
These encompass the naturalness of vehicle control, the workload imposed by the automated drive, the predictability of the vehicle's behavior, overall ride comfort, and the general driving style of the vehicle.
Participants' trust in the automated driving styles is assessed using the \ac{tia} questionnaire, originally composed of \num{19} items covering the reliability, predictability, familiarity, intention of developers, the propensity to trust, and trust in automation aspects.
Participants scored their agreement on each \ac{tia} item using a 5-point Likert scale, ranging from 1 (strongly disagree) to 5 (strongly agree).
Since the subjects had to answer these questions after each trip, we minimized the subjects' workload by selecting only the questions most relevant to this study.
To identify any potential instances of simulator sickness and to allow for the exclusion of affected participants from the statistical analysis when necessary, subjects filled out the \ac{ssq} \cite{kennedy1993simulator} at specific points during the study: shortly after training, at the halfway mark, and the study's end.


\begin{figure}[t]
    \centering
    \small
    \begin{tabular}{cc}
        \includegraphics[trim={0cm 0cm 0cm 0cm}, clip, width=.46\linewidth,valign=m]{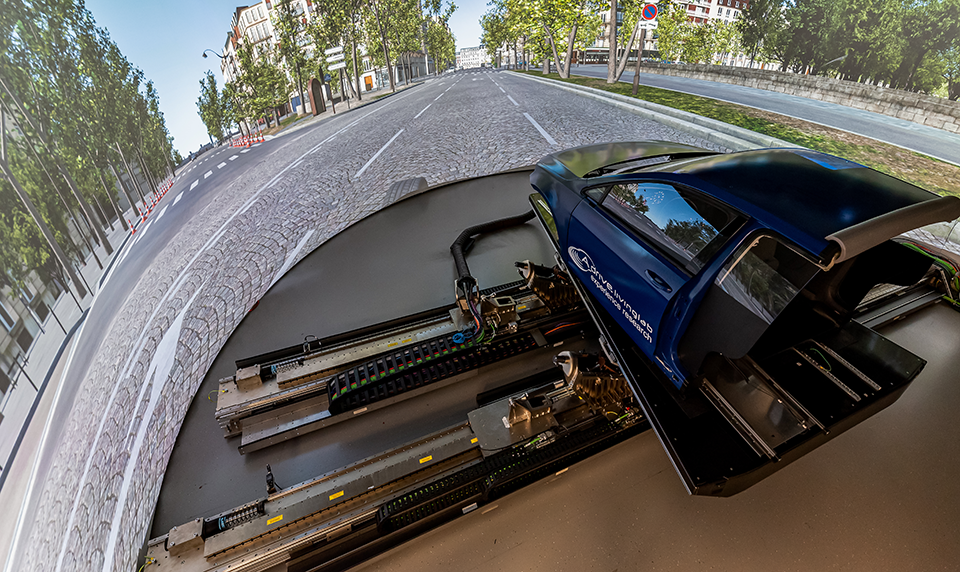} & 
        \includegraphics[trim={18.5cm 10cm 20.5cm 10cm}, clip, width=.46\linewidth,valign=m]{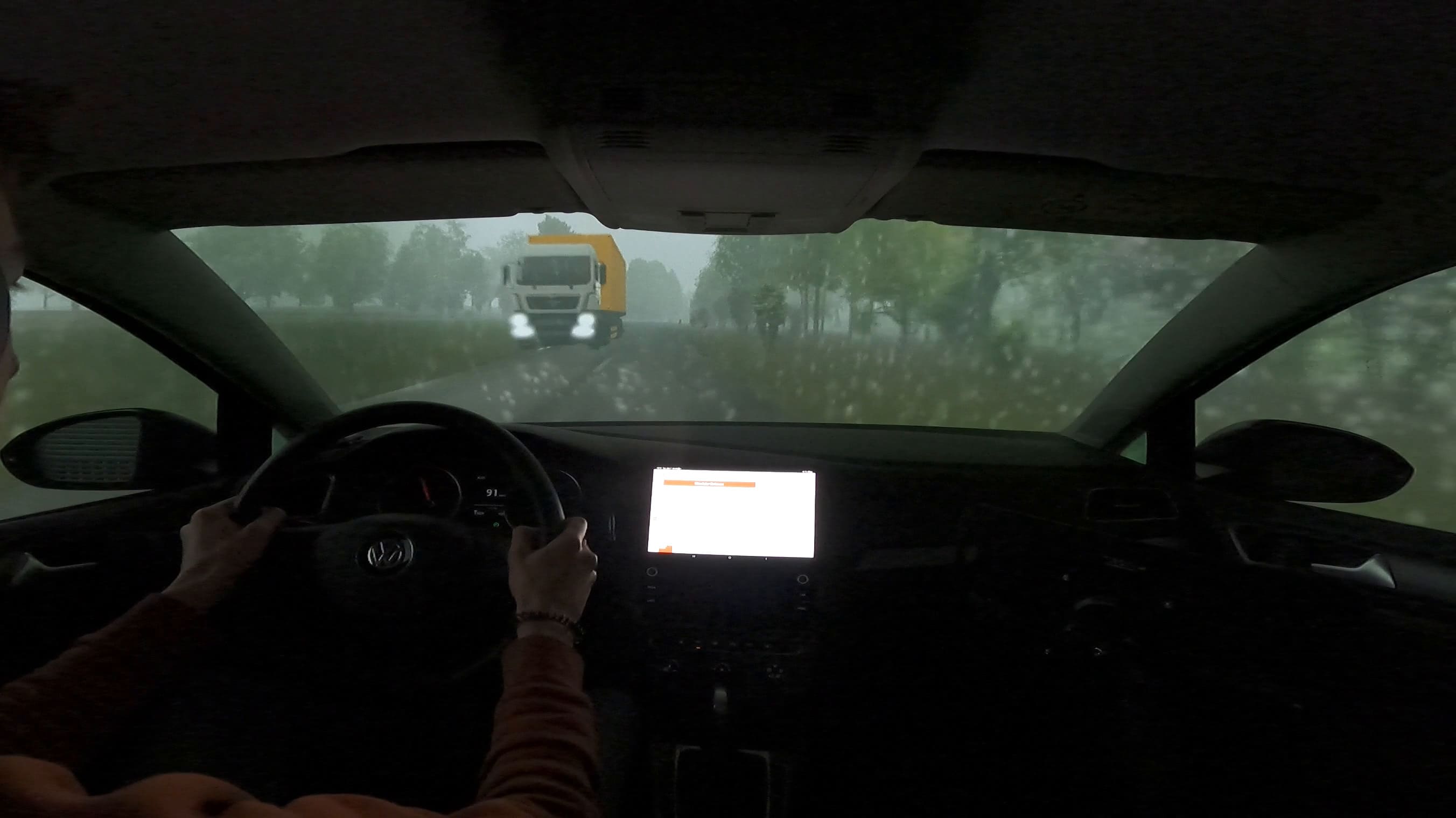} \\
        ~ & ~ \\
			a)  & b) \\
    \end{tabular}
    \caption{a) The Advanced Vehicle Driving Simulator features a six-degree-of-freedom motion platform, a series car cabin, and seven laser projectors. b) Adverse weather conditions in conjunction with oncoming traffic from the driver's perspective. A tablet positioned at the infotainment location is utilized to assess on-drive comfort.
    }
\label{fig:simEnv}
\end{figure}

\subsection{Simulator}
The virtual evaluation drives in this study were performed on the dynamic driving simulator at Kempten University of Applied Sciences.
The \ac{avds}, seen in \autoref{fig:simEnv}, features a six \ac{dof} motion platform driven by eight electric linear actuators, enabling the representation of vehicle motions with accelerations over \SI[per-mode = symbol]{10}{\metre\per\square\second} in all translational DOFs and over \SI[per-mode = symbol]{1100}{\degree\per\square\second} in all rotational DOFs.
Additional information regarding the static motion limits and dynamic platform performance can be found in \cite{kersten2022study}.
The \ac{avds} incorporates a Hardware-in-the-Loop steering test bench, allowing the simulation of a complete steering system with external rack force feedback \cite{schick2022use}.
An entire vehicle cabin represents the vehicle interior.
Seven laser projectors perform visualization of the simulation environment at a refresh rate of \SI{240}{\hertz} on a \SI{270}{\degree} cylindrical screen measuring eight meters in diameter at a height of four meters.
Seven dedicated rendering PCs receive the simulation output from a synchronization node at \SI{1}{\kilo\hertz}.
Vehicle, road, and tire models, as well as driver and lateral control models, are running on a real-time PC running RedHawk Linux, handling both parallel model execution and real-time IO communication via CAN, UDP, and EtherCAT at a sample rate of \SI{1}{\kilo\hertz}.
Platform motion is controlled by a separate real-time computer running the motion cueing algorithm as well as controllers for the platform actuators at \SI{2}{\kilo\hertz}.
Transmission of particularly time-critical model outputs and measurement feedback through a synchronized EtherCAT network is handled by a dedicated real-time PC oversampling at \SI{8}{\kilo\hertz}.

The virtual road is represented by a horizontal \SI{10}{\milli\metre} grid with a \SI{1}{\milli\metre} vertical resolution modeled from LiDAR data from the real-world reference road.
The tire is modeled using a classic Magic Formula 5.2 \cite{pacejka2005tire} parameter set that was parametrized using flat-track dyno measurements and validated using real measurements of the reference vehicle.
On the road model side, the contact patch is represented using a set of unweighted contact points calculated via the cylindrical surface of the nominal tire radius and width that intersects with the road surface model.
This intersection model returns only the averaged contact patch center position and normal vector to the tire model.
The virtual vehicle is simulated using a two-track, multibody model in IPG Carmaker, modeled to replicate a VW Golf VII GTD.
Wheels and chassis are modeled as rigid bodies, and the elastokinematic effects of the axles and wheels are represented by lookup tables based on ADAMS multi-body simulations.
The tire model is executed sequentially within the vehicle model simulation loop. It receives road data via the contact patch model, communicating with a dedicated terrain server running in cosimulation on the same real-time machine.

\subsection{Driving Styles}

\begin{figure}[]
    \centering
    \includegraphics[page=1,trim={0cm 0 0 0}, clip,width=1\linewidth]{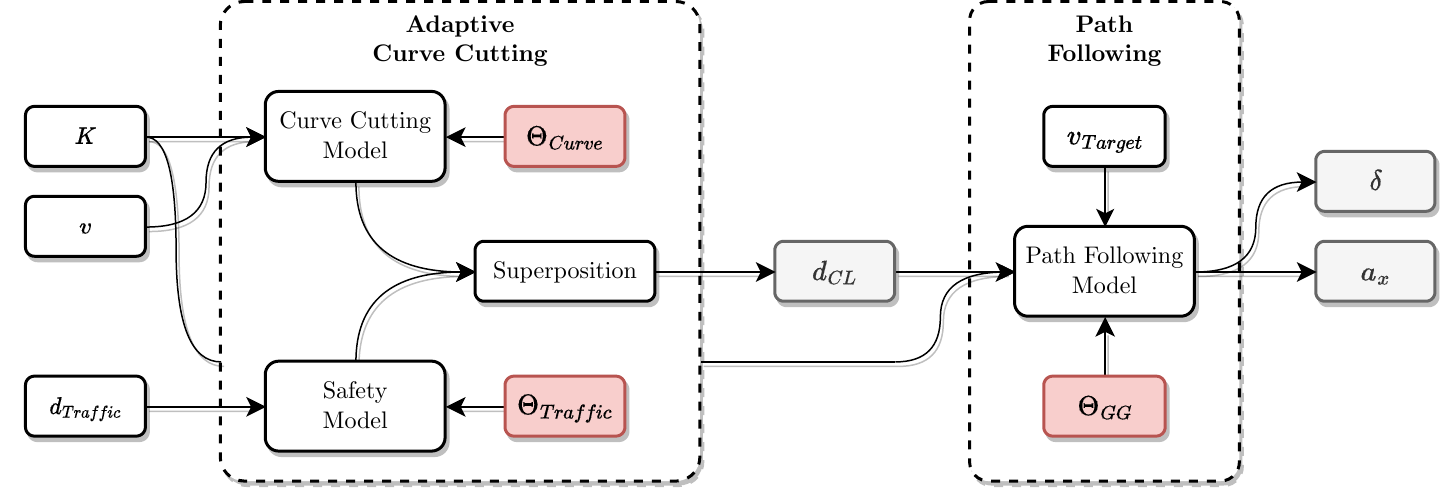}
    \caption{Overview of the cascaded longitudinal and lateral vehicle control function. Based on the curvature $K$, velocity $v$, and the longitudinal distance to the closest oncoming traffic object $d_{Traffic}$, the target distance to center lane value $d_{CL}$ is calculated. This intermediate result is utilized by the Path Following Model together with the target velocity $v_{Target}$ to calculate the steering wheel angle $\delta$ and the longitudinal acceleration $a_x$. The red-highlighted parameter vectors denote the variables that are adjusted to realize the distinct driving styles.}
    \label{fig:simDrivingStyles}
\end{figure}

The autonomous driving function is realized through a cascaded longitudinal and lateral vehicle controller consisting of an adaptive curve-cutting and path-following module. A high-level overview of the controller is shown in \autoref{fig:simDrivingStyles}.

\begin{figure*}[t]
    \centering
    \small
    \begin{tabular}{ccc}
        \includegraphics[trim={0cm 0cm -1.3cm 0cm}, clip,width=.43\linewidth,valign=m]{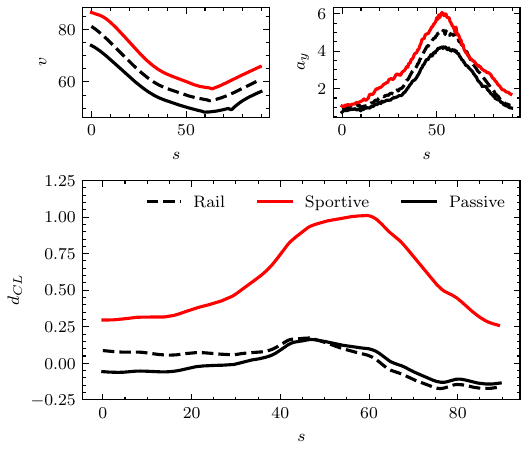} & 
        ~ &
        \includegraphics[trim={0cm 0cm -1.3cm 0cm}, clip,width=.43\linewidth,valign=m]{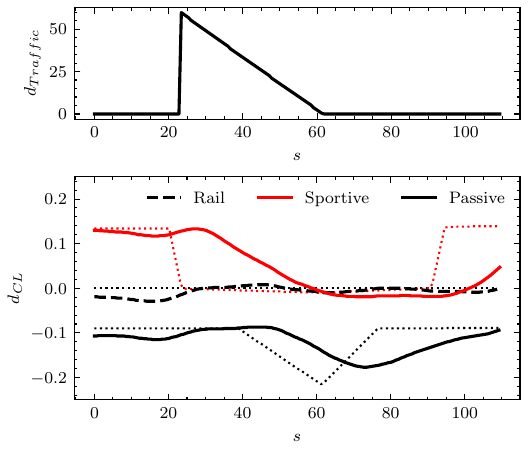} \\
			a) & ~ & b) \\
    \end{tabular}
    \caption{Comparison of the different cornering behaviors of the three driving styles. In a) the velocity $v$, lateral acceleration $a_y$, and the distance to the lane-center $d_{CL}$ are shown for a left curve without oncoming traffic.
    The sportive driving style shows the highest velocity, lateral acceleration, and curve-cutting values based on the parameterization.
    Constrained by the maximal acceleration values, the path-following model cannot hold the vehicle perfectly in the lane-center, which reduces the difference between the passive and rail driving styles in terms of curve cutting. In b), the reaction to an oncoming truck on a straight road segment is illustrated. The upper part of the figure shows the detected distance between the ego and the target vehicle. When the vehicle is detected, the passive and sportive driving style reduces their distance to the lane-center.
    In this case, the dashed lines represent the target values of the adaptive curve cutting module, and the solid lines represent the actual measured values achieved by the path after considering acceleration limits.}
\label{fig:simDriverModel}
\end{figure*}

\subsubsection*{Adaptive Curve Cutting}
The \ac{ccg} describes the stationary cornering behavior \cite{haselberger2023self,laubis2020ccg} and employs linear regression to calculate the gradient of the distance to the centerline $d_{CL}$ and the lateral acceleration $a_y$.
A positive $\mathrm{CCG}$ indicates curve cutting, whereas negative values suggest the vehicle drifting towards the curve's outside.
The intersection point with the y-axis determines the global offset $\mathrm{CCG}_0$.
Given its robust interpretability, the curve-cutting behavior is well-suited for industrial applications in developing driver assistance systems and autonomous driving functions \cite{barendswaard2019classification,hofer2020attribute}.

To mitigate the high-frequency fluctuations in the lateral acceleration signals, the lateral acceleration $a_y$ is calculated using the preview curvature $K$ and the current velocity $v$. The target distance to lane center $d_{CL}$ is calculated to:
\begin{equation}
    a_y = Kv^2 \\
\label{eq:dsAy}
\end{equation}
\begin{equation}
    d_{CL,CCG} = a_y CCG + CCG_0
\label{eq:dsdccg}
\end{equation}

where $d_{CL,CCG}$ is the curve-cutting proportion based on the human quasi-stationary curve-negotiation behavior.
Following \cite{din8855}, positive $d_{CL}$ values indicate a deviation towards the left, whereas negative values indicate a deviation towards the right lane boundary.
Drivers favor an early, noticeable action that aligns with the current situation \cite{lange2014automatisiertes}, aiming to address even minor potential risks as soon as possible \cite{bellem2018comfort}.
In this context, \cite{rossner2022also} recommends that autonomous vehicles should adapt their trajectory to the occurrence of oncoming traffic, as previous studies indicate that reactive trajectories lead to significantly higher acceptance, trust, and higher subjective driving experience \cite{rossner2018drive,rossner2019you,rossner2020does,rossner2020care,rossner2021hitting}.
In this study, the lateral shift induced by oncoming vehicles $d_{CL,T}$ is modeled as a linear relationship:
\begin{equation}
    \trafficGradient = \frac{d_{T,0}}{\trafficPreview}
\label{eq:nablaT}
\end{equation}
\begin{equation}
    d_{CL,T} = \min(\nabla_{T}d_{Traffic} + d_{T,0},0) 
\label{eq:dsdt}
\end{equation}

The gradient $\nabla_{T}$ signifies the rate at which this distance increases to the distance of the oncoming traffic $d_{Traffic}$ and is calculated based on the lateral traffic offset $\trafficLateralDistance$ and the longitudinal preview distance $\trafficPreview$.
The parameter $\trafficLateralDistance$ represents the lateral deviation of the ego vehicle from the center of the lane when both vehicles are exactly opposite each other.
To avoid sudden changes in the lateral offset, the gradient of $d_{CL,T}$ is limited to $0.5\nabla_T$:
\begin{equation}
    d_{CL,T,t-1}-0.5\nabla_T ~ \le ~ d_{CL,T,t} ~ \le ~ d_{CL,T,t-1}+0.5\nabla_T
\label{eq:dsdt_truncate}
\end{equation}
The final output results from the condition that the lateral displacement to the right, thus away from the oncoming traffic, must be at least $d_{CL,T}$:
\begin{equation}
    d_{CL} = \min(d_{CL,CCG},d_{CL,T}) 
\label{eq:dsdcl}
\end{equation}

\subsubsection*{Path Following}
To autonomously follow the road, a Path Following Model is needed. 
In this work, without loss of generalization, we employ the IPG Driver, a model specific to the used simulation environment.
By considering the lateral and longitudinal acceleration constraints $\Theta_{GG}$, the control parameters of steering angle and longitudinal acceleration are determined based on the desired driving velocity and the target distance to the lane center.
These constraints are modeled using a similar representation to the GG-Envelope defined in \cite{haselberger2023self}. 
Drivers with distinct driving styles display notable variations in lateral accelerations, indicating differences in their acceptable risk levels and experiences \cite{deng2020probabilistic}.
Acceleration values are frequently employed in the analysis of driving styles \cite{chen2019driving,vilaca2017systematic,kanarachos2018smartphones}.
Moreover, the resulting G-G diagrams depend on the driving style \cite{wegschweider2005modellbasierte,bae2020self} and evolve quickly while maintaining stability over time \cite{will2020methodological}.

\subsubsection*{Determinition of the Driving Style Parameters}
To simulate authentic driving behavior for the three autonomous driving styles passive, rail, and sportive, the specific parameters for the cascaded controller are extracted from real-world human driving on rural roads obtained from a comprehensive prior driving study \cite{haselberger2023self}. For passive, rail, and sportive, the 15$^{th}$ percentile, mean, and 85$^{th}$ percentile are used for the respective parameters and summarized in \autoref{tab:dsParams}.
Similar percentile-based approaches are applied in \cite{hajiseyedjavadi2022effect,wang2022classification,kamaraj2023comparing}.
The resulting different driving behaviors are exemplarily illustrated in \autoref{fig:simDriverModel}.
For all driving styles, a constant target speed is defined to exclusively account for the lateral variations between them.
However, the velocities experienced vary due to the acceleration constraints of the path-following module.
This implies that each driving style had its own resulting speed profile, which remained consistent across all participants and environmental conditions.
\subsection{Procedure}
Upon enrolling in the driving experiment, participants were provided with online questionnaires and requested to complete them before participating.
When participants arrived at the laboratory, they signed an institutionally approved consent form, including authorization to release anonymized data.
The introduction concluded with a general explanation of the study's objective, the dashboard evaluation application, and the overall simulator, withholding specific details about the various driving styles and scenarios to prevent any influence on the participants.
Subsequently, the participants were given \num{15} minutes to familiarize themselves with the simulator on a track different from the one used in the actual experiment.
A research assistant maintained continuous audio communication and guided the participants through the route. 
After completion of the training drive, the subject's condition was assessed using the \ac{ssq}, and if necessary, the experiment was interrupted.
For the main experiment, a $4 \times 2$ within-subjects design was applied. The first factor represents the four different driving styles passive, rail, sportive, and replay. As the second factor, all driving styles were evaluated in the two weather conditions clear and rainy.
The driving test was carried out on a digitized rural road featuring more than 30 curves and a five-kilometer length.
During each drive, the subjects encountered four oncoming trucks.
The scenarios were carefully scripted to guarantee that the trucks consistently approached the subjects at the same points on the track, irrespective of the exhibited driving style. 
Two of the four situations involving oncoming vehicles occur on left-hand curves, and the remaining two occur on straight segments of the road.
The curves, both with and without oncoming traffic, had similar radii to maintain comparability in the analysis.
Despite the demonstrated relative validity of high-fidelity simulators through consistent correlations with real-world driving scenarios \cite{helman2015validation}, simulator driving may exhibit heightened instability and occasional aggressiveness, potentially resulting in higher velocities compared to real-world conditions \cite{qi2019vehicle}.
To prevent result bias and isolate the impact of lateral driving behavior, the autonomous longitudinal control was consistently enabled while maintaining complete control of the vehicle steering.
Additionally, this ensures an alignment between the velocity profiles of manual and autonomous drives.

\begingroup

\setlength{\tabcolsep}{3pt} 

\begin{table}[]
    \caption{Parameters for the adaptive curve cutting and path following modules for the passive, rail, and sportive driving styles.}
    \centering
    \scriptsize
    \begin{tabular}{llcclcclccc}
        \toprule
    \multicolumn{1}{c}{\textbf{}} & \multicolumn{1}{c}{\textbf{}} & \multicolumn{2}{c}{\textbf{$\Theta_{Curve}$}} & \multicolumn{1}{c}{\textbf{}} & \multicolumn{2}{c}{\textbf{$\Theta_{Traffic}$}}            & \multicolumn{1}{c}{\textbf{}} & \multicolumn{3}{c}{\textbf{$\Theta_{GG}$}}                          \\ \cline{3-4} \cline{6-7} \cline{9-11}\clineSpacing 
    \multicolumn{2}{l}{\textbf{~}}                  & $CCG$     & $CCG_0$    & \textbf{}                     & $\trafficPreview$  & $d_{T,0}$ & \textbf{}                     & $\max(a_x)$ & $\min(a_x)$ & $\max(|a_y|)$  \\
    \multicolumn{2}{l}{\textbf{Driving Style}}      & \SI{}{\metre\per(\metre\per\square\second)}    & \SI{}{\metre}   & \textbf{}              & \unit{\metre}  & \SI{}{\metre} & \textbf{}                     & \SI{}{\metre\per\square\second} & \unit{\metre\per\square\second} & \SI{}{\metre\per\square\second}  \\
    \midrule
    Passive                       &                               & 0.042            & -0.09              &                               & 80                           & -0.218      &                               & 2.108          & -3.104         & 3.90                    \\
    Rail                          &                               & 0                & 0                  &                               & 80                            & 0           &                               & 3.206          & -3.916         & 4.731                  \\
    Sportive                      &                               & 0.141            & 0.135              &                               & 80                            & -0.01       &                               & 4.235          & -4.847         & 5.653                 \\
    \bottomrule
    \end{tabular}
\label{tab:dsParams}
\end{table}

\endgroup

After the training round, the subjects were asked to drive manually again under the pretext of getting to know the evaluation route under clear weather.
However, the participants' trajectories were secretly recorded for later playback.
Afterward, the participants were autonomously driven in a random sequence through four of the eight driving styles and weather conditions combinations.
An exception to this was the replay in rainy weather.
The subject's condition after the first half was assessed again using the SSQ to detect a possible occurrence of simulator sickness.
After a 20-minute break, the second part of the experiment resumed with a manual driving phase, this time under rainy conditions. Once more, the trajectories were covertly captured.
The remaining driving styles and weather combinations were presented to the subjects in the next step.
At the end of the simulator ride, the SSQ was filled out for the third time.
During each ride with an autonomous driving style, the subjects were asked to rate their subjective relaxation level.
Relying solely on a post-trial questionnaire could miss more nuanced responses, as immediate emotional reactions are easily forgotten \cite{ekman2019exploring}, while cognitive aspects tend to be more enduring \cite{norman2009way}.
An automated audio notification guaranteed that all participants provided ratings at the exact locations.
After each ride, participants completed the TiA and ARCA questionnaires using the tablet application.
The presented work follows The Code of Ethics outlined by the World Medical Association (Declaration of Helsinki) \cite{world2013world}.
Throughout the entire experiment, research assistants monitored the subjects through the audio link and strictly followed the safety regulations of the driving simulator, ensuring the participants' safety.


\section{Results}
The statistical analysis of the subjective self-evaluations and the objective driving indicators was performed in jamovi \cite{jamovi2023jamovi}, an open statistical software.
The \ac{tia}, \ac{arca}, and relaxation level questionnaires were analyzed with regard to a relationship between the autonomous driving style, and the driving context, consisting of the weather, road, and traffic situation.
Friedman Tests, non-parametric equivalents of Repeated Measures ANOVAs, were employed to assess whether there were statistically significant rating differences among the evaluated driving styles in the \ac{tia} and \ac{arca} questionnaires.
Jamovi's implementation of Durbin-Conover Post-hoc analyses \cite{pohlert2014pmcmr} was utilized to explore pairwise differences between the driving styles.
Paired samples T-tests were used to evaluate the significance of mean rating differences between the two weather conditions.
To identify significant group differences for the isolated analysis of weather, traffic, and curve effects on the on-drive comfort ratings, we employed independent samples t-tests and ANOVAs.
A summary of all responses split by the weather conditions and the AV's driving style can be seen in \autoref{fig:radarTiAARCA}, and the respective values can be found in \autoref{tab:inventoryResponses} in the Appendix.
Only the results that exhibited statistical significance ($p<.05$) are discussed in the following.

\begin{figure*}[t]
    \centering
    \small
    \begin{tabular}{cc}
        Dry & Rain \\
        ~ & ~ \\[-2mm]
        \includegraphics[trim={0cm 0.7cm 0cm 0cm}, clip, width=.33\linewidth,valign=m]{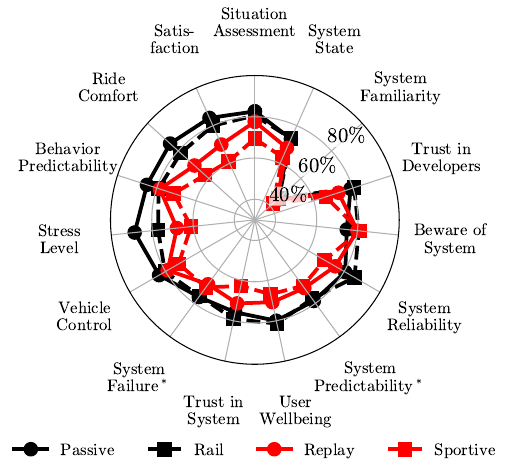}
         & 
            \includegraphics[trim={0cm 0.7cm 0cm 0cm}, clip, width=.33\linewidth,valign=m]{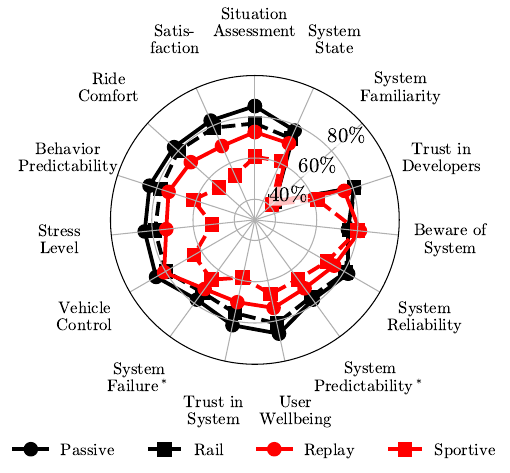}
        \\
        ~ & ~ \\
		\multicolumn{2}{c}{
            \includegraphics[trim={0cm 0.24cm 0cm 7.44cm}, clip, width=.40\linewidth,valign=m]{assets/after_ride_assessment_rain.pdf}
        } \\
    \end{tabular}
    \caption{Scaled mean ratings on the after-drive inventories \ac{tia} and \ac{arca} for the two weather conditions based on \autoref{tab:inventoryResponses}. 
    For a unified presentation, the items "Vehicle Control", "Stress Level", "Behavior Predictability", "Ride Comfort", and "Satisfaction" of the ARCA questionnaire are scaled using a maximum score of ten.
    The remaining items of the \ac{tia} inventory are scaled using a maximum score of five.
    The outermost circle represents the optimal achievable fulfillment of all criteria; inversed items are denoted with $^*$.
    }
\label{fig:radarTiAARCA}
\end{figure*}

\subsection{\acl{tia}}

The Cronbach's alpha for the \ac{tia} questionnaire was determined to be \num{0.907}, suggesting a high level of internal consistency reliability.
The ratings on the System Failure item showed significant differences between the AV's driving styles according to a Friedman Test \ChiSquarePL{3}{16.6}{.001}. 
Durbin-Conover pairwise comparisons revealed that the mean ratings were significantly lower for the passive \meanSD{2.15}{0.870} than for the sportive \meanSD{2.60}{1.065}, replay \meanSD{2.50}{1.027} and rail driving style \meanSD{2.28}{0.940}, \pValuePL.
In addition, mean ratings on the sportive driving style turned out to be significantly (\pValue{.025}) higher than for the lane center guidance.
Moreover, according to a Friedman Test \ChiSquarePL{3}{33.3}{.001}, the AV's driving style significantly impacts the ratings of the driving function's situation assessment capability.
Mean values were significantly higher for the passive driving style \meanSD{4.19}{0.715} than for the subjects' replay \meanSD{3.75}{0.992}, rail \meanSD{3.94}{0.871}, and the sportive driving style \meanSD{3.25}{0.926},  with \pValue{.003}, \pValue{.023}, and \pValuePL.
Durbin-Conover pairwise comparisons showed that mean ratings on the sportive driving style were significantly lower than for the rail and replay counterparts, \pValuePL. 
For the sportive driving style, mean ratings on situation assessment were found to be significantly higher in the clear \meanSD{3.47}{0.915} compared to rainy conditions \meanSD{3.03}{0.897}, as indicated by a Wilcoxon signed-rank test \wilcoxon{152.5}{.016}.
Additionally, the system state turned out to depend on the driving style. 
According to a Friedman Test \ChiSquarePL{3}{21.9}{.001}, the system state was significantly more transparent when driven by the passive \meanSD{3.76}{1.13} compared to the sportive automated driver \meanSD{3.11}{1.18} and the replayed trajectories \meanSD{3.47}{1.13} with \pValuePL~and \pValue{.005}.
The lane center guidance \meanSD{3.66}{1.14} was rated significantly (\pValuePL) better than sportive and significantly (\pValue{.022}) better than the drivers' replays.
Likewise, a Friedman Test \ChiSquarePL{3}{16.7}{.001} indicated significant mean differences regarding the system reliability.
The Post-hoc tests revealed that the ratings for both the passive \meanSD{4.07}{0.651} and the rail \meanSD{4.16}{0.570} driving style are significantly higher than for the sportive driving policy \meanSD{3.48}{0.755} with \pValue{.001}.
The lane center guidance was rated significantly (\pValue{.011}) more reliable than the trajectory replays \meanSD{3.74}{0.856}.

These findings are in line with the ratings on the trust level in the system \ChiSquarePL{3}{26.7}{.001} and trust towards the developers \ChiSquarePL{3}{28.3}{.001}, which also showed significant mean differences related to the driving style.
For the system trust, ratings were significantly higher for the passive \meanSD{3.95}{0.883} than for the sportive style \meanSD{3.04}{0.963} with \pValuePL ~and for the replay \meanSD{3.55}{1.008} with \pValue{.009}.
Rail \meanSD{3.89}{0.793} was rated significantly (\pValuePL) more trustworthy than the sportive driving policy.
Trust was also higher when driven by the own replay than the sportive driving style, \pValue{.002}.
Regarding the trust in the developers, ratings for the sportive design of the lateral control \meanSD{3.20}{1.29} were significantly lower compared to the passive control \meanSD{3.98}{1.08} and centered lane guidance \meanSD{4.03}{1.10} with \pValuePL.
Moreover, the Post-hoc tests revealed that the mean ratings on the trajectory replays \meanSD{3.70}{1.15} were significantly lower than for the passive and rail driving styles with \pValue{.002} and \pValue{.004}. 
However, the replays were rated significantly (\pValue{.028}) better than the sportive design of the lateral driving function.
Employing a Friedman Test \ChiSquarePL{3}{25.7}{.001}, similar mean difference patterns were identified concerning the perceived degree to which the developers take the users' well-being seriously.
Post-hoc tests revealed that ratings for the passive driver \meanSD{4.16}{1.06} were significantly (\pValuePL) higher compared to the replay \meanSD{3.61}{1.22} and sportive driver \meanSD{3.34}{1.29}.
Likewise, rail \meanSD{4.07}{1.10} was rated significantly (\pValuePL) higher than the replay and sportive one.

\begin{figure*}[t]
    \centering
    \small
    \begin{tabular}{ccc}
        \includegraphics[trim={0cm 0cm -1.3cm 0},width=.28\linewidth,valign=m]{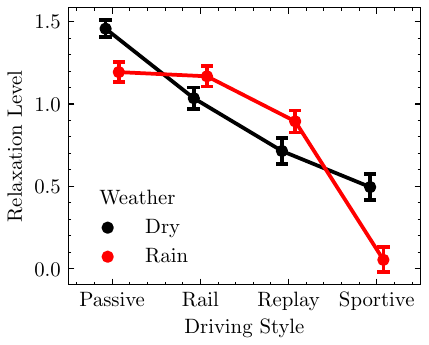} &
        \includegraphics[trim={0cm 0cm -1.1cm 0},width=.28\linewidth,valign=m]{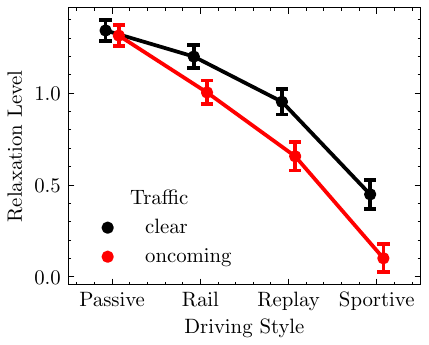} & 
        \includegraphics[trim={0cm 0cm -1.1cm 0},width=.28\linewidth,valign=m]{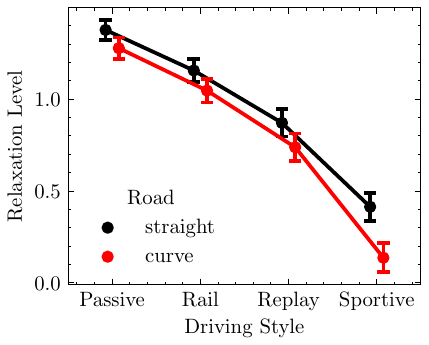} \\ 

    \end{tabular}
    \caption{
        Isolated evaluation of weather, traffic, and curve type effects on the on-drive subjective relaxation levels, numerical results can be found in \autoref{tab:relaxResponsesStats} in the Appendix.
        }
\label{fig:marginalMeansResults}
\end{figure*}

\subsection{\acl{arca}}
The Cronbach's alpha for the \ac{arca} questionnaire was \num{0.952}, indicating a high internal consistency reliability. 
In line with the evaluation of the \ac{tia}, similar user preferences towards the more passive behavior were also identified.
Significant mean differences in stress levels were observed for the various driving styles through a Friedman Test \ChiSquarePL{3}{45.3}{.001}.
The passive driving style \meanSD{8.62}{1.33} was perceived as significantly more relaxing than the other three presented driving styles rail \meanSD{7.79}{1.77}, replay \meanSD{7.05}{2.22}, and sportive \meanSD{5.60}{2.33} with \pValuePL.
Furthermore, the sportive driving style showed significantly (\pValuePL) higher stress values than rail and replay.
According to a paired samples T-Test $(\studentT{31.0}{3.061}{.005})$, the sportive driving style was significantly more relaxing under clear \meanSD{6.11}{2.32} than under rainy conditions \meanSD{5.09}{2.26}. 
The overall stress level results are also reflected in the satisfaction levels, which are significantly affected by the AV's driving style according to a Friedman Test \ChiSquarePL{3}{34.0}{.001}.
Satisfaction levels were significantly higher for the passive driving style \meanSD{8.34}{1.57} than for the drivers' replay \meanSD{6.96}{2.36}, \pValuePL.
The sportive driving style \meanSD{5.73}{2.34} was perceived as significantly less satisfactory than the passive, rail \meanSD{7.94}{1.46}, and replay counterparts, with \pValuePL, \pValuePL, and \pValue{.009}.

A Friedman Test \ChiSquarePL{3}{31.8}{.001} revealed significant mean differences among the driving styles in terms of the participants' ride comfort ratings.
The Post-hoc tests showed that the passive behavior \meanSD{8.40}{1.41} was significantly more comfortable than the drivers' replays \meanSD{7.06}{2.46} and rail \meanSD{7.92}{1.62} with \pValuePL~and \pValue{.024}.
The mean comfort ratings were also significantly lower for the sportive \meanSD{5.80}{2.40} compared to the passive, rail driving style, and the drivers' replay with \pValuePL, \pValuePL, and \pValue{.012}.
Following a Wilcoxon rank test \wilcoxon{234.0}{.016}, the sportive driving style was perceived as significantly more comfortable under clear \meanSD{6.25}{2.41} than under rainy conditions \meanSD{5.34}{2.34}. 
Driving style-specific mean differences were also found for the naturalness of the vehicle control \ChiSquarePL{3}{17.2}{.001} and the predictability of the vehicle's behavior \ChiSquarePL{3}{26.3}{.001}.
The passive variant \meanSD{8.44}{1.32} was rated significantly (\pValuePL) more natural than the sportive variant \meanSD{6.84}{2.07}.
Similarly, the sportive behavior was found to be significantly more unnatural than the lane center guidance \meanSD{8.01}{1.38} and the replayed trajectories \meanSD{7.98}{1.78} with \pValue{.023} and \pValuePL.
The naturalness of the sportive driving style revealed higher mean ratings for the clear weather situations \meanSD{7.27}{2.05} than for the scenario with rain \meanSD{6.41}{2.03}, \studentT{31}{2.698}{.011}.
In terms of the predictability of the vehicle's behavior, the sportive configuration \meanSD{6.64}{2.01} received significantly lower mean ratings than passive \meanSD{8.44}{1.34}, rail \meanSD{7.87}{1.75}, and replay \meanSD{7.63}{1.73} with \pValuePL, \pValue{.002}, and \pValuePL.
In addition, mean behavior predictability ratings of the passive driving style were significantly higher than for rail (\pValue{.006}) and replay (\pValue{.011}).
Regarding the weather conditions, significant mean differences were only found for the sportive driving style, \studentT{31}{2.625}{.013}. In dry situations \meanSD{7.14}{2.17}, the \ac{avs}' driving behavior was perceived as more predictable than in rainy situations \meanSD{6.14}{1.73}.

\subsection{On-Drive Stress Level Responses}
While the \ac{tia} and \ac{arca} questionnaires assess the overall impression after specific rides, the on-drive questionnaire on relaxation levels provides a direct means to examine the impact of the driving situation.
Irrespective of the driving style, a Yuen-Welch's test $(\yuenTPL{1222}{3.87}{.001})$ revealed that oncoming vehicles, in particular, had a significant adverse effect on user ratings.
Thereby, mean values for situations with oncoming vehicles \meanSD{0.767}{1.20} were significantly lower than those without vehicles on the opposing lane \meanSD{0.984}{1.12}.
Moreover, the differences between straight \meanSD{0.953}{1.13} and curvy \meanSD{0.798}{1.19} road sections are found to be statistically significant, \yuenT{1222}{2.59}{.010}.
A Mann-Whitney test $(\mannWhitney{489987}{.017})$ indicates that the significant mean stress level variations depended on the distinct weather conditions dry \meanSD{0.926}{1.17} and rain \meanSD{0.825}{1.15}.
In order to examine the driving style influences in isolation, the marginal mean relaxation levels for the four driving styles passive, rail, replay, and sportive are grouped by gender, weather, traffic, and road situation.
Mean differences were evaluated for significance using Mann-Whitney U tests and visualized in \autoref{fig:marginalMeansResults}.
All numerical results can be found in \autoref{tab:relaxResponsesStats} in the Appendix.
Across all splits, the overall order of driving style preferences regarding the relaxation levels stays consistent, with the passive being the most relaxed, followed by rail, replay, and sportive.
Regarding gender, it was found that men rated the passive driving style significantly better than women across all driving contexts.
This is reversed for the sportive driving style, where women provided significantly higher ratings than men.
The female participants also found the replay of their own trajectories more pleasant than the male test group, but no statistical significance was found for this difference.
Significant dependencies on weather were observed for the passive and sportive driving styles.
For the rain situation, the relaxation levels are, in each case, significantly lower for the more sportive design of the lateral vehicle control. 
Statistically not significant but noteworthy is the tendency for both rail and replay to receive better ratings in the rain than in dry conditions. 
Concerning the replay, this can be explained by the assumption that the subjects drove more carefully in adverse weather conditions, resulting in fewer instances of stress when faced with the recorded trajectories.
The most differences in relaxation levels were found for changing traffic situations.
When faced with oncoming vehicles, except for the passive driving style, all mean differences were significant, with the oncoming situation consistently rated more stressful.
The most significant deviation was found for the sportive driving behavior.
The almost identical ratings for the passive driving style were expected, as this style demonstrates the most pronounced reaction to oncoming traffic.
Although the curvy road sections were consistently rated more negatively than straight sections, the mean relaxation level differences only proved to be significant for the sportive driving style.

\begingroup

\setlength{\tabcolsep}{3pt} 

\begin{table}[t]
    \caption{
        Marginal mean relaxation levels for the four driving styles passive, rail, replay, and sportive grouped by weather, traffic, and road situation. Mean differences were evaluated for significance using Mann-Whitney U tests.}
    \centering
    \scriptsize
    \begin{tabular}{llcccc}
        \toprule
        \textbf{}       &   & \textbf{Passive}        & \textbf{Rail}           & \textbf{Replay}         & \textbf{Sportive}       \\
        \midrule
        Weather         & Dry             & 1.46 $\pm$ 0.83           & 1.04 $\pm$ 1.04           & 0.72 $\pm$ 1.29           & 0.50 $\pm$ 1.25          \\
                        & Rain            & 1.19 $\pm$ 0.93           & 1.17 $\pm$ 0.99           & 0.90 $\pm$ 1.07           & 0.06 $\pm$ 1.22          \\
                        &             & \pValuePL{}\pLevelThree  & \pValue{.159}  & \pValue{.574}  & \pValuePL{}\pLevelThree  \\
                        &                 &                         &                         &                         &                         \\
        Traffic         & Clear           & 1.34 $\pm$ 0.89           & 1.20 $\pm$ 0.99           & 0.95 $\pm$ 1.11           & 0.45 $\pm$ 1.25           \\
                        & Oncoming        & 1.31 $\pm$ 0.89           & 1.00 $\pm$ 1.04           & 0.66 $\pm$ 1.24           & 0.10 $\pm$ 1.23           \\
                        &             & \pValue{.646}  & \pValue{.022}\pLevelOne & \pValue{.013}\pLevelOne & \pValue{.002}\pLevelTwo \\
                        &                 &                         &                         &                         &                         \\
        Road            & Straight        & 1.38 $\pm$ 0.85           & 1.16 $\pm$ 1.01            & 0.87 $\pm$ 1.18           & 0.41 $\pm$ 1.22           \\
                        & Curve           & 1.28 $\pm$ 0.93           & 1.05 $\pm$ 1.03            & 0.74 $\pm$ 1.19           & 0.14 $\pm$ 1.27           \\
                        &             & \pValue{.244} & \pValue{.190} & \pValue{.172} & \pValue{.015}\pLevelOne \\
                        \bottomrule
        \multicolumn{6}{l}{\footnotesize{Note: \pLevelOne~$p<.05$, \pLevelTwo~$p<.01$, \pLevelThree~$p<.001$}} \\
        \end{tabular}
    
\label{tab:relaxResponsesStats}
\end{table}

\endgroup

\subsection{Subjective Driving Style Classification}
After each drive, the subjects classified the experienced driving style.
The clearest overlap can be found in the sportive driving style, with a total of \num{79.9} percent.
The passive and replay driving styles are recognized similarly well, with \num{45.3} and \num{48.4} percent.
However, the driving style that always keeps the vehicle in the center of the lane turns out to be the most diffuse.
Only \num{34.4} percent of the subjects correctly classified this driving style.

\begingroup

\setlength{\tabcolsep}{1.7pt} 

\begin{table}[h]
    \caption{Significant ($p<.01$) correlations between the MDSI factor scores and the questionnaire items queried during and after the simulator drives.}
    \centering
    \scriptsize
    \begin{tabular}{lllllc}
        \toprule
        \textbf{Factor Score} & \textbf{} & \textbf{Inventory}  & \textbf{Item}            & \textbf{}      & \textbf{Correlation} \\
        \midrule
        Angry                   &           & \ac{arca}             & Ride Comfort             &                & -0.221\pLevelThree             \\
                                &           &                       & Satisfaction             &                & -0.211\pLevelThree               \\
                                &           &                       & Stress Level             &                & -0.211\pLevelThree               \\
                                &           & On-Drive              & Relaxation Level         &                & -0.136\pLevelThree               \\
                                &           &                       &                          &                &                      \\
        Anxious                 &           & \ac{arca}             & Stress Level             &                & -0.198\pLevelTwo               \\
                                &           &                       & Vehicle Control          &                & -0.225\pLevelThree               \\
                                &           & On-Drive              & Relaxation Level         &                & -0.111\pLevelTwo               \\
                                &           & \ac{tia}              & Situation Assessment     &                & -0.164\pLevelTwo               \\
                                &           &                       & System Failure           &                & 0.284\pLevelThree                \\
                                &           &                       & System Familiarity       &                & 0.164\pLevelTwo                \\
                                &           &                       & System Predictability    &                & 0.238\pLevelThree                \\
                                &           &                       & System Reliability       &                & -0.224\pLevelThree               \\
                                &           &                       & Trust in System          &                & -0.236\pLevelThree               \\
                                &           &                       &                          &                &                      \\
        Careful                 &           & \ac{arca}             & Behavior Predictability  &                & 0.185\pLevelTwo                \\
                                &           & \ac{tia}              & User Wellbeing           &                & 0.276\pLevelThree                \\
                                &           &                       & Situation Assessment     &                & 0.190\pLevelTwo                \\
                                &           &                       & System Predictability    &                & -0.175\pLevelTwo               \\
                                &           &                       & System Reliability       &                & 0.239\pLevelThree                \\
                                &           &                       &                          &                &                      \\
        Dissociative            &           & \ac{tia}              & User Wellbeing           &                & -0.149\pLevelTwo               \\
                                &           &                       &                          &                &                      \\
        Risky                   &           & \ac{tia}              & User Wellbeing           &                & -0.280\pLevelThree               \\
                                &           &                       & System Familiarity       &                & -0.235\pLevelThree               \\
                                &           &                       & System Predictability    &                & 0.189\pLevelTwo                \\
        \bottomrule
        \multicolumn{6}{l}{\footnotesize{Note: Partial Pearson correlations, two-tailed test,}} \\
        \multicolumn{6}{l}{\footnotesize{controlling for age and gender}} \\
        \multicolumn{6}{l}{\footnotesize{\pLevelTwo~$p<.01$, \pLevelThree~$p<.001$}} \\
        \end{tabular}
\label{tab:mdsiCorrelations}
\end{table}

\endgroup
\subsection{Correlation of Subjects' Driving Style and AV Driving Style Preference}
To compare the self-assessment of the subjects' driving styles with the ratings on the \ac{arca}, \ac{tia}, and the on-drive relaxation questionnaires in the first step, the factor scores are calculated based on the MDSI-DE questionnaire.
Given the limited number of subjects in this study, which is considered insufficient for a substantial factor analysis of the MDSI items, similar to \cite{van2018relation,haselberger2023self} the factor divisions and loadings are adopted from \cite{van2015measuring} and to maximize the validity of the estimates, multiple regressions are utilized to calculate refined scores \cite{van2015measuring, karjanto2017identification, nees2021relationships,distefano2009understanding}.
To verify that the effects can be attributed entirely to the driving style scores, partial correlations were computed controlling for age and gender, similar to \cite{van2018relation,holman2015romanian}.
The significance threshold was set to $p < .01$.
Results are summarized in \autoref{tab:mdsiCorrelations} in the Appendix.

Overall, only small effect sizes occurred.
The most correlations were found for the anxious factor. Participants who scored high on this factor showed significantly lower relaxation values on the \ac{arca} and the on-drive measures.
There was also a negative correlation regarding the system's reliability, trust in the system, situation assessment, and the naturalness of vehicle control.
Furthermore, individuals categorized under this driving style generally gave higher ratings to the likelihood of a system failure.
All of this suggests that individuals with higher anxiety levels were more prone to harbor concerns about the automated driving function.
The anxious factor score positively correlates with system familiarity and system predictability, suggesting that these subjects are more acquainted with similar systems and believe they can anticipate their behavior.
The strongest negative correlation regarding on-drive relaxation levels was found with the angry factor score.
Regardless of the autonomous vehicle's driving style, weather, and traffic situation, the general relaxation level, satisfaction, and ride comfort values were consistently rated lower.
Also, there is a tendency to reject the automated driving function in this case, even if no significant correlations with the actual comfort indicators can be found.
A comparable trend is evident in the risky factor score, which exhibits a significant negative correlation with the perceived degree to which developers prioritize passenger well-being.
Although participants with high scores on this factor reported significantly lower familiarity with similar systems, the system predictability correlates positively with this score.
In contrast, the most positive correlations were found for the careful factor score.
Individuals scoring high on this factor demonstrated significantly elevated ratings in behavior predictability, user well-being, situation assessment, and overall system reliability.
However, the system's ability to accurately assess situations was given lower overall ratings.
No significant correlations were found regarding the perceived comfort.
In the dissociative factor score, only the perceived attention of developers to passenger well-being exhibits a significant negative correlation. In contrast, for the distress-reduction score, no significant correlations were identified.
\section{Discussion}

This study aimed to investigate the differences in driving style preferences on rural roads under different traffic and weather conditions using subjective trust and comfort ratings.
Furthermore, we aimed to assess whether driving style information obtained from self-reports correlates with trust and comfort evaluations, providing insights into connecting participants' personal driving styles with their preferences for an AV's driving style.
Overall, it is evident that participants exhibited a distinct preference for the passive driving style.
This driving style is characterized by reduced curve-cutting gradients, lower accelerations, and a more noticeable response to oncoming traffic.
This aligns with numerous studies, including \cite{wang2022classification,peng2022drivers,yusof2016exploration,ma2021drivers,basu2017you,hartwich2015drive,bellem2018comfort,sourelli2023user,rossner2020you,ekman2019exploring,dillen2020keep}, indicating that users generally favor a more passive driving style when being driven by an AV.
On the contrary, in previous works, participants generally rated the aggressive driving style the lowest, irrespective of their mood \cite{phinnemore2021happy} and personal traits \cite{bellem2018comfort}.
This trend is also evident in the current study, with the sportive driving style consistently receiving lower trust, comfort, and relaxation ratings.

The assessment of the overall impression following the specific rides revealed significantly higher predictability values for the passive compared to the sportive driving style, consistent with findings from prior studies \cite{ekman2019exploring}.
Predictability is considered a critical factor in building trust at the early stages of an interaction \cite{rempel1985trust}.
It is also an integral component of the performance information required to sustain an appropriate level of trust \cite{lee2004trust}.
Accordingly, the passive driving style was perceived as the most trustworthy.
The test subjects' lack of trust in the system of the sportive lateral guidance is also evident in the elevated system failure ratings.
Passengers prefer a passive autonomous driving style with lower speeds, smoother accelerations, and earlier brakings \cite{beggiato2020komfopilot}.
Trust and acceptance of automated driving systems also depend on the perceived safety \cite{molnar2018understanding,natarajan2022toward,lee2004trust,dixit2019risk,detjen2020wizard}.
According to \cite{ma2021drivers}, aggressive AVs are more commonly taken over due to discomfort, perceived safety concerns, or anxiety, as the driving styles of aggressive AVs exceed drivers' safety margins \cite{summala2007towards}.
The observation within \cite{price2016psychophysics, hajiseyedjavadi2022effect} that accurate lane-center tracking was considered more competent than less precise tracking by the subjects further explains this study's lower favored curve-cutting values.

In this context, the level of situational assessment plays a crucial role in shaping trust towards an AV \cite{petersen2019situational}.
Interestingly, the test subjects assigned significantly lower situation awareness ratings to the replay of their own driving style than to the passive driving style.
Likewise, the scores for overall trust, satisfaction, and comfort were significantly lower for the replay compared to the passive driving style.
In this domain, earlier research findings are inconclusive, with some studies suggesting that subjects prefer their own driving style \cite{hajiseyedjavadi2022effect,sun2020exploring,karlsson2021encoding,griesche2016should}, while others indicate the opposite \cite{basu2017you,scherer2016will}.
The results in \cite{scherer2015driver} and \cite{hartwich2015drive} indicate that the driving style preference was age-dependent.
The disparity between the manual driving style and the preference for an AV can be explained by the perception that the driving dynamics encountered in conditional automated driving are perceived as more demanding than those experienced during manual driving \cite{vasile2023influences}.
Subjects generally favor lower speeds when they are not manually driving \cite{horswill1999effect}.
Furthermore, divergences between subjective and objective evaluations indicate that certain drivers perceive their driving as non-aggressive, contrary to objective measures \cite{sarwar2017grouped}.

There is high evidence in the literature that the driving context significantly influences subjective evaluations of driving style \cite{dillen2020keep,vasile2023influences,oliveira2019driving, hajiseyedjavadi2022effect, beggiato2019physiological,radhakrishnan2020measuring,peng2022drivers,hartwich2018driving,rossner2019diskomfort,ossig2022tactical}, as drivers' preferences are not fixed and change based on their state and the situation \cite{hasenjager2019survey,yi2019implicit,angkititrakul2009evaluation,lin2014overview}.
Regarding the on-drive relaxation level responses, the weather conditions significantly negatively affected the passive and sportive driving styles.
Previous studies have also reported differing levels of subjective comfort based on changing environmental conditions \cite{siebert2019speed,hajiseyedjavadi2022effect}.
While not statistically significant, the subjects' trajectory replay results showed an interesting pattern, as the mean ratings were higher for the rainy conditions.
A plausible explanation for this phenomenon could be attributed to the heightened vigilance and cautious driving behavior exhibited in these situations.
This is substantiated by the observation that drivers generally exhibit increased caution during rainy conditions \cite{hamdar2016weather}.
The level of perceived situational awareness plays a crucial role in shaping trust towards an AV \cite{petersen2019situational}.
This demonstrates that a weather-independent design of the automated driving style may lead to a more significant reduction in comfort and a diminished sense of safety under adverse weather conditions.

The assessment of on-drive evaluations revealed that the presence of oncoming traffic reduced relaxation values.
There were significant deteriorations except for the passive driving style with the maximum observed in the sportive driving style.
These findings align with the results in \cite{rossner2020care}, where perceived safety was found to be influenced by the position, type, and quantity of oncoming traffic.
Trucks elicit significantly greater perceived safety concerns compared to cars.
The absence of a significant decline in comfort for the passive driving style can be attributed to its distinct response to oncoming traffic.
Similarly, in \cite{rossner2020care}, reacting to oncoming traffic results in a notably higher perception of safety.
This more human-like behavior contributes to a higher perceived anthropomorphism and can positively influence how automated systems are perceived \cite{waytz2014mind, musabini2021park4u, ruijten2018enhancing}.
Conversely, if the driving style of the automated vehicle is not adaptive to oncoming vehicles and the resulting lane positioning is perceived as inadequate for the current situation, it diminishes the user's trust \cite{lee2016question}.

High evidence in the literature suggests a strong correlation between driving style preferences, personality traits, and individual driving styles \cite{peng2022drivers, ellinghaus2001beifahrer,bruck2021investigation,bellem2018comfort,ma2021drivers, hajiseyedjavadi2022effect,louw2019relationship,yusof2016exploration,delmas2022effects}.
The findings of this study also indicate that distinct driver types perceive various automated driving styles in disparate ways.
Similar to \cite{ma2021drivers}, the impact of the AV's driving style on the subjects' ratings is more significant for defensive drivers than for aggressive drivers.
Notably, drivers scoring high on the Anxious factor exhibited significantly lower on-drive relaxation values and overall lower \ac{tia} and \ac{arca} ratings.
This aligns with the findings in \cite{hajiseyedjavadi2022effect}, where low sensation-seeking individuals feel discomfort with more sportive controllers.

The results in this study demonstrate that the prevailing assumption in the literature, suggesting that drivers prefer a driving style that mirrors their own \cite{hasenjager2019survey,festner2016einfluss,griesche2016should,bolduc2019multimodel,sun2020exploring,hartwich2015drive,rossner2022also,dettmann2021comfort}, only holds for some subjects, with the majority showing preferences towards the traffic adaptive passive driving style.
Other studies have also indicated that not all subjects necessarily prefer an exact mimicking of their driving style \cite{hartwich2018driving} and some individuals may favor more conservative driving styles \cite{basu2017you,scherer2016will,bellem2018comfort,yusof2016exploration,dillen2020keep,beggiato2020komfopilot,peng2022drivers}.

The assessment of subjective driving style classification after each individual ride reveals difficulties for the subjects in differentiating between the various driving styles.
In the final survey after the study, it was identified that \SI{47}{\percent} of the test subjects clearly perceived general differences between the driving styles, \SI{50}{\percent} perceived them only partially, and \SI{3}{\percent} did not perceive them at all.
The different curve-cutting behaviors were perceived as clearly distinguishable by \SI{56}{\percent} of the test subjects and as partially distinguishable by the remaining \SI{44}{\percent}.
The reaction to oncoming traffic was not consciously perceived by all test subjects.
Only \SI{15}{\percent} were able to detect apparent differences, while the majority of participants (\SI{65}{\percent}) reported only partial differences.
Twenty percent could not detect any differences.
Contrasting this with the driving style-specific parameters and the resulting significantly different distances to oncoming traffic, it becomes evident that the test subjects perceive these trajectory adjustments subconsciously rather than consciously.
That the design of the driving style regarding their reaction to oncoming traffic nonetheless has a measurable effect is evident from the fact that significant differences between the driving styles can be observed in the actual evaluation of the relaxation level.
Only \SI{18}{\percent} distinctly recognized their trajectory replay and \SI{23}{\percent} were unable to identify any similarities. The remaining \SI{59}{\percent} of the test subjects recognized at least partial overlaps.
In general, this aligns with the findings in \cite{basu2017you}, which concludes that most drivers are unable to classify their own driving style, even into broad behavioral clusters, or identify their style.

Subjects not only misclassified their driving style but also assigned it mediocre ratings, which were generally significantly lower compared to the passive and rail driving styles.
This supports the assumption in \cite{basu2017you} that drivers prefer to experience automated driving in a manner they believe aligns with their own driving style, irrespective of their actual driving style.
Moreover, users' assessment of an AV's driving style is shaped by a combination of objective and subjective factors \cite{peng2022drivers}.
Individuals commonly use their personal driving behavior as a reference point when evaluating different driving styles \cite{rossner2019diskomfort}.
However, there is a chance that driving style preferences change when self-driving cars become more widely used \cite{phinnemore2021happy}.
With more trust in the technology, the users can become comfortable with more aggressive driving styles.

\section{Limitations}
A potential limitation of our study is that drivers were observed in an unfamiliar environment.
Merely \num{29.4}\%  of the participants indicated prior or regular experience with driving a simulator.
We attempted to mitigate this effect by using a series production cabin and letting the subjects experience the simulator and the vehicle during the familiarization phase without giving them specific instructions.
Additionally, it is important to note that driving simulators cannot fully replicate real-world conditions \cite{blana2002differences, godley2002driving,groeger2020driver}, potentially resulting in differences between the driving behavior observed in simulator experiments and that in real-world driving \cite{xue2019rapid,wang2017driving,corcoba2021covid,van2018relation}.
Some drivers perceive situations within a simulator as less hazardous in contrast to real-world situations and drive more aggressively \cite{corcoba2021covid}, reasoning that no one will get injured \cite{helman2015validation}.
However, there is substantiated evidence that a driving simulator serves as a valid tool for analyzing driving behavior \cite{van2018relation, changbin2015driving, zhao2014partial, meuleners2015validation,schluter2021identifikation}.

Research involving instrumented vehicles or simulators can make participants conscious of their involvement in an experiment, potentially introducing biases compared to real-world driving \cite{carsten2013vehicle, shrestha2017hardware}.
Despite our efforts, like clearly informing drivers that the study was not an exam, the potential for observer bias to impact the results still exists.
However, in specific studies \cite{quimby1999drivers,grayson2003risk}, the impact of this effect is considered to be less substantial than initially assumed.
Another limitation of this study is the exclusive use of two extremes: predefined driving styles or exact replication of the subjects' trajectories.
However, it is reasonable to assume that combinations of predefined styles or an adaptation to the demonstrated driving behavior could influence the preferences and evaluations of the participants.
With a total duration of two hours for the entire execution of all driving scenarios, it did not seem productive to evaluate more driving style variations.
Therefore, this remains a subject for future work.

In this study, self-reported driving styles were examined among German drivers.
While the MDSI has been used in various countries, its factorial structure may vary slightly.
The sample size is considered insufficient for a robust factor analysis of the MDSI items.
Therefore, factor divisions and loadings are adopted from a previous study \cite{van2015measuring}.
However, differences in subjects' country-specific attributes between the two studies may limit generalizability.
One further limitation of the experiment is that the current study solely involved a subjective on-drive assessment instead of objective physiological measures like galvanic skin response or heart rate. Analyzing data across various levels of aggregation can enhance the reliability of findings.
Moreover, drawing conclusions from studies with small sample sizes and uneven gender distribution, as in this case, requires caution.
Sampling errors may lead to overestimating the magnitude of any genuine observed effects.
Furthermore, limited statistical power can present difficulties in initially identifying such effects \cite{button2013power}.
\section{Conclusion}
Research on autonomous vehicles aims to reduce traffic accidents.
However, the widespread use of this technology depends on the comfort experience of the passengers.
In this work, a controlled driving study was conducted using a high-fidelity driving simulator to assess subjects' preferences for different driving styles, focusing on being driven on rural roads under various traffic and weather conditions.
We propose a reactive driving behavior model capable of emulating human-like curve negotiation while responding to oncoming traffic to simulate different driving behaviors and safety margins.
Statistical analyses of participants' responses during and after the drives unveiled a distinct inclination toward the more passive driving style, characterized by a low curve-cutting gradient, moderate lateral and longitudinal acceleration constraints, and a pronounced reaction to oncoming traffic.
The assumption that the test subjects prefer to be driven by mimicking their own driving behavior could not be confirmed in this study.
In fact, participants rated their trajectory replay significantly lower than the more general passive driving style.
Moreover, it was demonstrated that both weather conditions and oncoming traffic significantly influence perceived relaxation and trust values during autonomous rides.
Additionally, the individual driving style of the test subjects, as represented by the six driving style factors derived from the MDSI, was found to impact perceived well-being and sense of safety significantly when being driven by autonomous driving styles.

We recommend that future research focus on incorporating a more detailed representation of the driving context beyond ego vehicle-dependent quantities into the personalization process of AV's driving styles.
The results indicate a high correlation between changing driving situations and the experienced passenger comfort.
To facilitate further research, we made the dataset publicly available.

\section*{LIST OF ABBREVIATIONS}

\begin{acronym}\itemsep-10pt
	\acro{ads}[ADS]{Aggressive Driving Scale}
	\acro{aic}[AIC]{Akaike Information Criteria}
	\acro{arca}[ARCA]{Automated Ride Comfort Assessment}
	\acro{avds}[aVDS]{Advanced Vehicle Driving Simulator}
	\acro{avs}[AVs]{Autonomous Vehicles} 
	\acro{ccg}[CCG]{Curve Cutting Gradient}
	\acro{csai}[CSAI-2]{Competitive State Anxiety Inventory 2}
	\acro{dbi}[DBI]{Driving Behaviour Inventory}
	\acro{dbq}[DBQ]{Driver Behavior Questionnaire}
	\acro{dof}[DOF]{Degree of Freedom}
	\acro{dscf}[DSCF]{Dwass-Steel-Critchlow-Fligner} 
	\acro{dsq}[DSQ]{Driving Style Questionnaire}
	\acro{fms}[FMS]{Fast Motion Sickness Scale}
	\acro{loc}[LoC]{Locus of Control}
	\acro{mdsi}[MDSI]{Multidimensional Driving Style Inventory}
	\acro{rcnn}[RCNN]{Recurrent Convolutional Neural Network}
	\acro{ssq}[SSQ]{Simulator Sickness Questionnaire}
	\acro{tas}[TAS]{Thrill and Adventure Seeking}
	\acro{tia}[TiA]{Trust in Automated Systems}

\end{acronym}



\section*{CREDIT AUTHORSHIP CONTRIBUTION STATEMENT}
\textbf{Johann Haselberger:} Conceptualization, Methodology, Software, Validation, Formal analysis, Investigation, Writing - original draft, and Writing - Review \& Editing. \newline
\textbf{Maximilian Böhle:} Software, Writing - original draft, and Writing - Review \& Editing \newline
\textbf{Bernhard Schick:} Conceptualization and Writing - Review \& Editing. \newline
\textbf{Steffen Müller:} Conceptualization and Writing - Review \& Editing. 
\section*{ACKNOWLEDGMENTS}
The authors extend their gratitude to Stefanie Trunzer, Mike Köhler, and Philipp Rupp from the Institute for Driver Assistance and Connected Mobility, University of Applied Science Kempten, for their invaluable assistance with the driving simulator.
Additionally, the authors want to thank Marco Eichberger for his
contributions during the subject management and test execution. 
\section*{DATA AVAILABILITY}
The dataset including the anonymized soci-demographics and questionnaire responses, the raw vehicle measurements including labels, and the derived driving style indicators is publicly available.
\section*{CONFLICTS OF INTEREST}
The authors declare no conflict of interest.

\newpage
\appendix
\section{Appendix}

\begingroup

\setlength{\tabcolsep}{2.5pt} 

\begin{table}[h!]
    \caption{Mean and standard deviation results of the \ac{tia}, \ac{arca}, and on-drive relaxation level questionnaires split by the weather condition and the AV's driving style.}
    \centering
    \scriptsize
    \begin{tabular}{llllcccclcccc}
        \toprule
        \textbf{}           & \textbf{}            & \textbf{}      & \textbf{} & \multicolumn{4}{c}{\textbf{Dry}}                                       & \textbf{} & \multicolumn{4}{c}{\textbf{Rain}}                                      \\ \cline{5-8} \cline{10-13}\clineSpacing 
        \textbf{Inventory}  & \textbf{Item}        & \textbf{Style} & \textbf{} & \rotatebox{90}{\textbf{Passive}} & \rotatebox{90}{\textbf{Rail}} & \rotatebox{90}{\textbf{Replay}} & \rotatebox{90}{\textbf{Sportive}} & \textbf{} & \rotatebox{90}{\textbf{Passive}} & \rotatebox{90}{\textbf{Rail}} & \rotatebox{90}{\textbf{Replay}} & \rotatebox{90}{\textbf{Sportive}} \\
        \midrule
        \acs{tia}           & Situation                      & Mean           &           & 4.13             & 4.03          & 3.88            & 3.47              &           & 4.26             & 3.84          & 3.63            & 3.03              \\
                            & Assessment                     & SD             &           & 0.83             & 0.93          & 0.91            & 0.92              &           & 0.58             & 0.81          & 1.07            & 0.90              \\[\tableVSpacing]
                            & System                         & Mean           &           & 3.65             & 3.66          & 3.41            & 3.16              &           & 3.87             & 3.66          & 3.53            & 3.06              \\
                            &    State                       & SD             &           & 1.25             & 1.26          & 1.21            & 1.14              &           & 0.99             & 1.04          & 1.05            & 1.24              \\[\tableVSpacing]
                            & System                         & Mean           &           & 2.22             & 2.28          & 2.22            & 2.09              &           & 2.16             & 2.16          & 2.19            & 2.06              \\
                            &    Familiarity                 & SD             &           & 1.34             & 1.28          & 1.18            & 1.33              &           & 1.21             & 1.35          & 1.35            & 1.08              \\[\tableVSpacing]
                            & Trust in                       & Mean           &           & 3.94             & 4.03          & 3.63            & 3.31              &           & 4.03             & 4.03          & 3.78            & 3.09              \\
                            &  Developers                    & SD             &           & 1.11             & 1.18          & 1.18            & 1.23              &           & 1.08             & 1.03          & 1.13            & 1.35              \\[\tableVSpacing]
                            & Beware of                      & Mean           &           & 3.74             & 4.03          & 4.00            & 4.06              &           & 3.77             & 3.81          & 4.06            & 4.00              \\
                            &   System                       & SD             &           & 1.18             & 0.93          & 0.92            & 0.98              &           & 1.02             & 1.20          & 0.91            & 0.98              \\[\tableVSpacing]
                            & System                         & Mean           &           & 4.03             & 4.29          & 3.75            & 3.45              &           & 4.11             & 4.04          & 3.72            & 3.52              \\
                            &  Reliability                   & SD             &           & 0.68             & 0.54          & 0.93            & 0.78              &           & 0.63             & 0.59          & 0.80            & 0.74              \\[\tableVSpacing]
                            & System                         & Mean           &           & 2.07             & 2.14          & 2.46            & 2.52              &           & 2.11             & 2.19          & 2.45            & 2.72              \\
                            & Predictability                 & SD             &           & 1.07             & 1.15          & 1.26            & 0.95              &           & 1.13             & 0.74          & 0.91            & 0.80              \\[\tableVSpacing]
                            & User                           & Mean           &           & 4.00             & 4.07          & 3.54            & 3.34              &           & 4.32             & 4.07          & 3.68            & 3.34              \\
                            &    Wellbeing                   & SD             &           & 1.09             & 1.14          & 1.29            & 1.23              &           & 1.02             & 1.07          & 1.16            & 1.37              \\[\tableVSpacing]
                            & Trust in                       & Mean           &           & 3.79             & 3.96          & 3.57            & 3.14              &           & 4.11             & 3.81          & 3.54            & 2.93              \\
                            &    System                      & SD             &           & 1.00             & 0.81          & 0.96            & 0.99              &           & 0.74             & 0.79          & 1.07            & 0.94              \\[\tableVSpacing]
                            & System                         & Mean           &           & 2.19             & 2.22          & 2.57            & 2.48              &           & 2.11             & 2.33          & 2.43            & 2.71              \\
                            &     Failure                    & SD             &           & 0.83             & 1.05          & 1.10            & 0.98              &           & 0.92             & 0.83          & 0.96            & 1.15              \\
                            &                                &                &           &                  &               &                 &                   &           &                  &               &                 &                   \\
        \acs{arca}          & Vehicle                        & Mean           &           & 8.36             & 8.03          & 7.88            & 7.27              &           & 8.53             & 7.98          & 8.08            & 6.41              \\
                            &      Control                   & SD             &           & 1.31             & 1.51          & 1.98            & 2.05              &           & 1.36             & 1.24          & 1.55            & 2.03              \\[\tableVSpacing]
                            & Stress                         & Mean           &           & 8.86             & 7.72          & 6.80            & 6.11              &           & 8.37             & 7.87          & 7.32            & 5.09              \\
                            &     Level                      & SD             &           & 1.21             & 2.19          & 2.53            & 2.32              &           & 1.43             & 1.23          & 1.85            & 2.26              \\[\tableVSpacing]
                            & Behavior                       & Mean           &           & 8.50             & 7.91          & 7.84            & 7.14              &           & 8.37             & 7.82          & 7.40            & 6.14              \\
                            &     Predictability             & SD             &           & 1.34             & 2.02          & 1.61            & 2.17              &           & 1.35             & 1.44          & 1.85            & 1.73              \\[\tableVSpacing]
                            & Ride                           & Mean           &           & 8.53             & 7.84          & 6.94            & 6.25              &           & 8.27             & 8.00          & 7.18            & 5.34              \\
                            &    Comfort                     & SD             &           & 1.46             & 1.92          & 2.65            & 2.41              &           & 1.36             & 1.26          & 2.28            & 2.34              \\[\tableVSpacing]
                            & Satisfaction                   & Mean           &           & 8.41             & 8.00          & 7.00            & 6.11              &           & 8.27             & 7.89          & 6.92            & 5.36              \\
                            &                                & SD             &           & 1.60             & 1.67          & 2.57            & 2.27              &           & 1.55             & 1.25          & 2.21            & 2.39              \\
                            &                                &                &           &                  &               &                 &                   &           &                  &               &                 &                   \\
        On Drive            & Relaxation                     & Mean           &           & 1.46             & 1.04          & 0.71            & 0.50              &           & 1.19             & 1.17          & 0.89            & 0.05              \\
                            &      Level                     & SD             &           & 0.83             & 1.04          & 1.29            & 1.25              &           & 0.93             & 0.99          & 1.07            & 1.22              \\
                            \bottomrule
        \end{tabular}
\label{tab:inventoryResponses}
\end{table}

\endgroup





\bibliographystyle{elsarticle-harv} 
\bibliography{References}






\end{document}